\documentclass[preprint,12pt,authoryear]{elsarticle}

\usepackage{amssymb}
\usepackage{graphicx}%
\usepackage{multirow}%
\usepackage{amsmath,amssymb,amsfonts}%
\usepackage{amsthm}%
\usepackage{mathrsfs}%
\usepackage[title]{appendix}%
\usepackage{xcolor}%
\usepackage{textcomp}%
\usepackage{manyfoot}%
\usepackage{booktabs}%
\usepackage{algorithm}%
\usepackage{algorithmicx}%
\usepackage{algpseudocode}%
\usepackage{listings}%
\usepackage{lscape}
\usepackage{setspace}
\usepackage{lipsum}
\usepackage{url}
\usepackage{tikz}
\usetikzlibrary{patterns}
\usepackage{multicol}
\usepackage{xcolor}

\journal{}

\begin{document}

\setlength{\tabcolsep}{2pt}
\begin{frontmatter}



\newtheorem{theorem}{Theorem}
\newtheorem{assumption}{Assumption}
\newtheorem{proposition}{Proposition}
\newtheorem{corollary}{Corollary}
\newtheorem{lemma}{Lemma}
\newtheorem{remark}{Remark}

\title{Spatial disaggregation of time series}


\author[label1,label3,label4]{A. Tobar}
\author[label1,label2,label3,label4]{A. Mir}
\author[label1,label2,label3,label4]{R. Alberich}
\author[label1,label2,label3,label4]{I. Garcia-Mosquera}
\author[label1,label3,label4]{M. Miró}
\author[label1,label2,label3,label4]{NA. Cruz\corref{cor1}}

\cortext[cor1]{Universitat de les Illes Balears, Departament de Matemàtiques i Informàtica, Phone: +34 637 54 6888, Palma de Mallorca, España, nelson-alirio.cruz@uib.es}
\address[label1]{Artificial Intelligence Research Institute of the Balearic Islands (IAIB), Department of Mathematics and Computer Science, University of the Balearic Islands, Palma 07122, Spain}
\address[label2]{Health Research Institute of the Balearic Islands (IdISBa), Palma 07010, Spain}
\address[label3]{Laboratory of Artificial Intelligence Applications (LAIA@UIB), Department of Mathematics and Computer Science, University of the Balearic Islands, Palma 07122, Spain}
\address[label4]{Data Modelling and Statistical Learning (MoDAE), Department of Mathematics and Computer Science, University of the Balearic Islands, Palma 07122, Spain}

\begin{abstract}
Spatiotemporal modeling of economic aggregates is increasingly relevant in regional science due to the presence of both spatial spillovers and temporal dynamics. Traditional temporal disaggregation methods, such as Chow-Lin, often ignore spatial dependence, potentially losing important regional information. We propose a novel methodology for spatiotemporal disaggregation, integrating spatial autoregressive models, benchmarking restrictions, and auxiliary covariates. The approach accommodates partially observed regional data through an anchoring mechanism, ensuring consistency with known aggregates while reducing prediction variance. We establish identifiability and asymptotic normality of the estimator under general conditions, including non-Gaussian and heteroskedastic residuals. Extensive simulations confirm the method’s robustness across a wide range of spatial autocorrelations and covariate informativeness. The methodology is illustrated by disaggregating Spanish GDP into 17 autonomous communities from 2002 to 2023, using auxiliary indicators and principal component analysis for dimensionality reduction. This framework extends classical temporal disaggregation to the spatial domain, providing accurate regional estimates while accounting for spatial spillovers and irregular data availability.
\end{abstract}


\end{frontmatter}

\section{Introduction}

Spatiotemporal data modeling has gained increasing relevance in econometrics and regional science due to the simultaneous presence of spatial spillovers and temporal dynamics in key macroeconomic indicators. Spatial autoregressive (SAR) models have long been used to capture spatial autocorrelation between neighboring regions, improving inference and prediction by explicitly considering spatial dependencies \cite{lesage2015introduction, cressie2011statistics}. Recent developments have extended these models to incorporate temporal structure, leading to comprehensive spatiotemporal frameworks that model both forms of dependence jointly \cite{gelfand2019handbook, hsu2021matrix}.

A particularly relevant application of this framework is the temporal disaggregation of macroeconomic aggregates, such as GDP, across subnational units. In many countries, national statistical agencies publish annual regional GDP figures, while only national totals are available on a quarterly basis. For policy evaluation, business cycle analysis, and regional monitoring, there is a strong demand for regionally consistent annual or quarterly estimates. Benchmarking techniques, such as the Chow-Lin method \cite{chow1971best} and its Eurostat-approved variants \cite{eurostat2013benchmarking}, have been widely applied for temporal disaggregation using high-frequency auxiliary indicators. However, these approaches do not incorporate spatial dependence between regions, potentially missing important information.

Recent studies have begun to explore the integration of spatial elements into disaggregation and forecasting, including machine learning and autoregressive alternatives (\citet{cuevas2015}, \citet{han2024sarn}), but they often lack a complete econometric or interpretable framework. In this context, it is desirable to develop models that maintain the inferential transparency of classical methods while considering spatial dependence and the irregular availability of regional data.

This article contributes to the literature by proposing a new methodology for the spatiotemporal disaggregation of economic aggregates, combining spatial autoregressive models, benchmarking restrictions, and covariate information. Before presenting our approach, it is important to highlight several features that make it attractive both theoretically and practically. First, the method is based on a rigorous asymptotic theory: we establish global identifiability and asymptotic normality of the joint estimator.
Second, we introduce a novel anchoring mechanism that incorporates partially observed regional values as linear restrictions. This step mimics real-life situations with sporadic regional data (e.g., for capital cities or large communities) and helps reduce the variance of the predictions. Third, we conducted an exhaustive simulation study encompassing 74,088 parameter configurations, covering a wide range of spatial autocorrelations, covariate informativeness, and grid sizes. These simulations confirmed the robustness and stability of our approach across various scenarios. Finally, we demonstrated the practical relevance of our method with a case study on the disaggregation of Spanish GDP into the 17 autonomous communities between 2002 and 2023, using a large set of auxiliary indicators and principal component analysis for dimensionality reduction.

Thus, our methodology extends the classical Chow-Lin approach to the spatial domain, enabling flexible and accurate estimation of regional GDP, consistent with national aggregates, while considering both spatial spillovers and the availability of partial observations.

\section{Spatial autoregressive model}
\subsection{Definition of the model}
Let $Y_t$ be a time series measured at time $t=1, \ldots, T$ over the spatial unit $\mathcal{R}$, which is subdivided in $n$ spatial regions $\{\mathcal{R}_1,\ldots ,\mathcal{R}_n\}$; let $\mathbf{z}_{it}$ a set of $k$ known explicative variables measured at time $t$ in spatial region $i$. Let $Y_{it}$ be the $i$th unknown time serie disagregated at time $t$ over the region $i$ with the restriction defined as:
\begin{equation}
    \sum_{i=1}^n Y_{it} = Y_t,\qquad \forall t=1, \ldots, T.
\end{equation}
The objective is to obtain  estimated values for $Y_{it}$ using the information allocated both in $\mathbf{z}_{it}$ and $\mathbf{Y}_a=(Y_1, \ldots, Y_T)$. Let $\pmb{\beta}=(\beta_1, \beta_2, \ldots, \beta_{k})^\top$ be the associate vector of parameters effects of $\mathbf{z}_{it}$ over $\mathbf{Y}$, a spatial autoregressive model is supposed for the vector $\mathbf{Y}_{t}=(Y_{1t}, \ldots, Y_{nt})$,  the completed model is given by:
\begin{align}
    &Y_{it}=\rho \sum_{j=1}^n w_{ij}Y_{jt} + \mathbf{z}_{it}^\top\pmb{\beta} + u_{it}. \label{sart}
\end{align}
Let
\begin{align}\label{align:model_data}
    \mathbf{Z}&=\begin{bmatrix}
    \mathbf{z}_{11}^\top\\
    \mathbf{z}_{21}^\top\\
    \vdots\\
    \mathbf{z}_{n1}^\top\\
    \mathbf{z}_{12}^\top\\
    \vdots \\
    \mathbf{z}_{n2}^\top\\
    \vdots\\
    \mathbf{z}_{nT}^\top\\
    \end{bmatrix}_{nT\times k},\; 
    \mathbf{Y}=\begin{bmatrix}
    Y_{11}\\
    Y_{21}\\
    \vdots\\
    Y_{n1}\\
    Y_{12}\\
    \vdots \\
    Y_{n2}\\
    \vdots\\
    Y_{nT}\\
    \end{bmatrix}_{nT\times 1},\; 
    \mathbf{W}=\begin{bmatrix}
    w_{11}& \cdots & w_{1n}\\
    \vdots & \ddots & \vdots\\
    w_{n1} & \cdots & w_{nn}\\
    \end{bmatrix},\; \mathbf{U}=\begin{bmatrix}
    u_{11}\\
    u_{21}\\
    \vdots\\
    u_{n1}\\
    u_{12}\\
    \vdots \\
    u_{n2}\\
    \vdots\\
    u_{nT}\\
    \end{bmatrix}_{nT\times 1}.
\end{align}
and $ \mathbf{A}=\mathbb{I}_{nT}-\rho (\mathbb{I}_{T}\otimes \mathbf{W})=\mathbb{I}_T\otimes (\mathbb{I}_{n}-\rho  \mathbf{W})$ and $\mathbb{I}_{T}$ the $T$-dimensional identity matrix. Then, the model defined in Equation \eqref{sart} can be expressed as:
\begin{align}
\mathbf{Y} &= \rho (\mathbb{I}_{T}\otimes \mathbf{W}) \mathbf{Y} + \mathbf{Z}\pmb{\beta} + \mathbf{U}\nonumber\\
\mathbf{Y} -\rho (\mathbb{I}_{T}\otimes \mathbf{W}) \mathbf{Y} &= \mathbf{Z}\pmb{\beta} + \mathbf{U}\nonumber\\
   (\mathbb{I}_{nT}-\rho (\mathbb{I}_{T}\otimes \mathbf{W}))\mathbf{Y} &= \mathbf{Z}\pmb{\beta} + \mathbf{U}\nonumber\\
    \mathbf{Y} &= \mathbf{A}^{-1}\mathbf{Z}\pmb{\beta} +\mathbf{A}^{-1} \mathbf{U}.\label{sar2}
 \end{align}
Let $\mathbf{U}_i=(u_{i1}, \ldots, u_{iT})$ be an $ARIMA(1,0,0)$, i.e. $\phi_1(B) u_{it} = \epsilon_{it}$ 
This proposal models the unobservable time series analogously to the proposal of \citet{proietti2006temporal}, \citet{quilis2018temporal}, and \citet{bisio2024reconstructing}. This assumption over $\mathbf{U}_i$ allows us to define the covariance matrix of $\mathbf{U}_i$ like \cite{10208005}, that is:
\begin{equation}
    \pmb{\Sigma}_U=Var(\mathbf{U}_i)=\frac{\sigma^2}{1-\phi_1^2}\begin{pmatrix}
        1 & \phi_1 & \phi_1^2 &\cdots & \phi_1^{T-1}\\
        \phi_1 & 1 & \phi_1 &\cdots & \phi_1^{T-1}\\
        \phi_1^2 & \phi_1 & 1 &\cdots & \phi_1^{T-1}\\
        \vdots & \vdots & \vdots &\ddots & \vdots\\
         \phi_1^{T-1}& \phi_1^{T-2} &\phi_1^{T-3} & \cdots & 1\\
    \end{pmatrix}.\label{eq:sigmaU}
\end{equation}
where $\phi_1$ is the autoregresive parameter of the $ARIMA(1,0,0)$ and $\sigma^2$ their associated variance. The problem is that $\mathbf{Y}$ and $\mathbf{U}$ are unobservable by the characteristics of the measurements. But $\mathbf{Z}$ and $\mathbf{Y}_a$ are observable and known, then, we define the matrix $\mathbf{C}$ as:
\begin{equation}
    \mathbf{C}= \mathbb{I}_{T}\otimes \mathbf{1}_{n}^\top,
\end{equation}
where $\mathbf{1}_{n}^\top$ is a row matrix of ones, such that
 \begin{align}
    \mathbf{CY} &= \mathbf{Y}_a =\mathbf{C}\mathbf{A}^{-1} \mathbf{Z}\pmb{\beta} + \mathbf{C}\mathbf{A}^{-1}\mathbf{U}.\label{sar3}
\end{align}
We obtain that:
\begin{align}
    \mbox{Var}(\mathbf{U})_{nT\times nT}&= \pmb{\Sigma}_U\otimes \mathbb{I}_n, \nonumber \\
    \mbox{Var}(\mathbf{C}\mathbf{A}^{-1}\mathbf{U})_{T\times T} &= \mathbf{C}\mathbf{A}^{-1} (\pmb{\Sigma}_U\otimes \mathbb{I}_n) (\mathbf{A}^{-1})^{\top}\mathbf{C}^\top = \mathbf{C}\mathbf{B}\mathbf{C}^\top=\pmb{\Sigma}_{\mathbf{Y}_a}\label{normY},\\
    \mbox{Cov}(\mathbf{Y}, \mathbf{Y}_a)_{nT\times T} &=  \mbox{Cov}(\mathbf{Y}, \mathbf{CY})= \mbox{Var}(\mathbf{Y}, \mathbf{Y})\mathbf{C}^\top  = \mathbf{B}\mathbf{C}^\top,\label{normYYa}
\end{align}
where $\mathbf{B}=\mathbf{A}^{-1} (\pmb{\Sigma}_U\otimes \mathbb{I}_n) (\mathbf{A}^{-1})^{\top}=\pmb{\Sigma}_{\mathbf{Y}}$. 
\subsection{Estimation and Inference}
To develop the maximum likelihood theory, we will first make assumptions of normality and homoscedasticity regarding the model innovations.
\begin{assumption}\label{as1}
 The innovations satisfy $\epsilon_{it}\sim \mathcal{N}(0, \sigma^2)$ for all $i=1, \ldots,n$ and $t=1,\ldots,T$.
\end{assumption}
Under Assumption \ref{as1}, properties of normal distribution (\citet{harvey1990forecasting}, \citet{durbin2012time}), and Equations \eqref{normY} and \eqref{normYYa} it follows that
\begin{align}
\mathbf{U}&\sim \mathcal{N}_{nT}(\mathbf{0}, \pmb{\Sigma}_U\otimes \mathbb{I}_n),\nonumber\\
\mathbf{Y}&\sim \mathcal{N}_{nT}(\mathbf{A}^{-1} \mathbf{Z}\pmb{\beta}, \mathbf{B}),\nonumber\\
\mathbf{Y}_a &\sim \mathcal{N}_{T}(\mathbf{C}\mathbf{A}^{-1} \mathbf{Z}\pmb{\beta}, \pmb{\Sigma}_{\mathbf{Y}_a}),\label{NormalAgreg}\\
\begin{pmatrix}
    \mathbf{Y}\\
    \mathbf{Y}_a
\end{pmatrix}&\sim \mathcal{N}_{nT+T}\left(\begin{pmatrix}
    \mathbf{A}^{-1} \mathbf{Z}\pmb{\beta}\\
    \mathbf{C}\mathbf{A}^{-1} \mathbf{Z}\pmb{\beta}
\end{pmatrix}, \begin{pmatrix}
    \mathbf{B} & \mathbf{BC}^\top,\\
    \mathbf{C}\mathbf{B} & \pmb{\Sigma}_{\mathbf{Y}_a}
\end{pmatrix}\right),\label{normalBig}\\
\mathbf{Y}\mid \mathbf{Y}_a &\sim \mathcal{N}_{nT}(\pmb{\mu}+\mathbf{BC}^\top\pmb{\Sigma}_{\mathbf{Y}_a}^{-1}(\mathbf{Y}_a-\pmb{\mu}_a), \mathbf{B}-\mathbf{BC}^\top\pmb{\Sigma}_{\mathbf{Y}_a}^{-1}\mathbf{CB}),\label{CondNormal}
\end{align}
where $\pmb{\mu}= \mathbf{A}^{-1} \mathbf{Z} \pmb{\beta}$, and $\pmb{\mu}_a = \mathbf{C}\pmb{\mu}$.
Therefore, by Equation \eqref{NormalAgreg},  the generalized least squares estimator for $\pmb{\beta}$ is given by:
\begin{align}
    \hat{\pmb{\beta}}&= (\mathbf{Z}^\top (\mathbf{A}^{-1})^{\top} \mathbf{C}^\top\pmb{\Sigma}_{\mathbf{Y}_a}^{-1}\mathbf{C}\mathbf{A}^{-1} \mathbf{Z})^{-1}\mathbf{Z}^\top (\mathbf{A}^{-1})^{\top} \mathbf{C}^\top\pmb{\Sigma}_{\mathbf{Y}_a}^{-1}\mathbf{Y}_a.\label{betasar}
\end{align}
It should be noted that the estimator $\hat{\pmb{\beta}}$ depends on the values of the parameters $\rho$, $\sigma^2$, and $\phi_1$. To estimate these parameters, from Equation \eqref{NormalAgreg}, we have that the density function of $\mathbf{Y}_a$ is given by:
\begin{equation*}
    f_{\mathbf{Y}_a}(\mathbf{y})=(2\pi)^{-\frac{T}{2}}\vert\pmb{\Sigma}_{\mathbf{Y}_a} \vert^{-\frac{1}{2}}\exp \left(-\frac{1}{2}(\mathbf{Y}_a - \pmb{\mu}_a)^\top \pmb{\Sigma}_{\mathbf{Y}_a}^{-1} (\mathbf{Y}_a - \pmb{\mu}_a)\right), \text{ for }\mathbf{y}\in\mathbb{R}^{T},
\end{equation*}
then, we solve the maximization problem over the log-likelihood function
\begin{equation}
  (\hat{\rho}, \hat{\sigma}^2, \hat{\phi}_1)
= \underset{(\rho, \phi_1, \sigma^2) \in \mathcal{U}}{\mbox{argmax}}
\Bigg\{
- \frac{1}{2} \log |\pmb{\Sigma}_{\mathbf{Y}_a}|
- \frac{1}{2} (\mathbf{Y}_a - \pmb{\mu}_a)^\top \pmb{\Sigma}_{\mathbf{Y}_a}^{-1} (\mathbf{Y}_a - \pmb{\mu}_a)
\Bigg\}\label{rhosar}  
\end{equation}
where $\mathcal{U}=(-1,1)\times(-1,1)\times(0,\infty)$. 

In practice, solving the problem of Equation \eqref{rhosar} can be done using \texttt{optim} function from software  \cite{Rmanual} with the "L-BFGS-B" algorithm, which allows the constraints on the parameter space $\mathcal{U}$ to be explicitly imposed. The L-BFGS-B algorithm is a limited-memory version of the Broyden–Fletcher–Goldfarb–Shanno (BFGS) quasi-Newton method, adapted for large-scale problems with simple box constraints \cite{byrd1995limited}. 

To ensure that the parameters are estimable and estimators are sufficient, consistent, and asymptotically normal, the following conditions are proposed:
\begin{assumption}\label{as2}
    The elements $w_{ij}$ satisfy that $\sum_{j=1}^n w_{ij}=1$ for all $i=1, \ldots,n$. And exist at least one $j\neq j'$ such that:
    \begin{equation*}
        \sum_{i=1}^n w_{ij} \neq \sum_{i=1}^n w_{ij'}  
    \end{equation*}
\end{assumption}
Assumption~\ref{as2} ensures that at least one spatial unit has a difference between what its neighbors contribute to it and what it contributes to its neighbors in spatial information. This prevents the degenerate case where regions are indistinguishable from one another, yielding multiple solutions.
Additionally, the matrix is standardized by rows to ensure that the parameter $\rho$ falls between -1 and 1, facilitating a more accurate interpretation of the autoregressive parameter. This condition is discussed usually in SAR models \citep{lee2004asymptotic, kelejian2010specification}
\begin{assumption}\label{as3}
  The number of spatial regions $n>2$ is fixed for all $t=1, \ldots, T$, $\vert \rho \vert<1$ and, $\vert \phi_1 \vert<1$.
\end{assumption}
Assumption~\ref{as3} ensures a minimum number of spatial regions and bounds $\rho$ and $\phi_1$ to guarantee the stationarity of the ARIMA process. 

\begin{assumption}\label{as345}
The matrix $(\mathbf{C}\mathbf{A}^{-1}\mathbf{Z}:\mathbf{C}\mathbf{A}^{-1}(\mathbb{I}_T\otimes\mathbf{W}) \mathbf{A}^{-1}\mathbf{Z} \pmb{\beta})$ has full column rank, that is, the $k+1$ columns are linearly independent.
\end{assumption}
Assumption~\ref{as345} is a natural translation of Assumption 8 of \citet{lee2004asymptotic} in the context of our model. This assumption ensures that both $\pmb{\beta}$ and $\rho$ are identifiable. Notice that $T>k+1$ is assumed. 

This condition is generally satisfied when the number of time periods $T$ greatly exceeds the number of parameters $k+1$, i.e., $T>>k+1$. Intuitively, each time block provides an independent observation of the effect of the spatial operator embedded in $\mathbf{A}^{-1}$ acting on the regressors $\mathbf{Z}_t$.

\begin{assumption}\label{as4}
    The elements of $\mathbf{Z}$ are uniformly bounded constants for all $n$ and $T$. The $\lim_{T\to \infty} \frac{1}{T}\mathbf{Z}^\top \mathbf{Z}$ exists and is nonsingular.
\end{assumption}
Assumption \ref{as4} enforces  $\mathbf{Z}$ to be full column rank. This assumption guarantees $\pmb{\beta}$ to be well-behaved as a regression estimator in Equation \eqref{betasar}. If this assumption is not met, decomposition techniques such as PCA or PLS, can be performed to guarantee an alternative $\tilde{\pmb{Z}}$ matrix that has full column rank.

We propose the following iterative procedure to obtain estimators of $(\pmb{\beta}^\top, \rho, \sigma^2, \phi_1)$ and the disaggregated spatio-temporal series:

\begin{itemize}
    \item[Step 1. ] Compute $\hat{\rho}, \hat{\phi}_1, \hat{\sigma}^2$ of Equation \eqref{rhosar}, using \texttt{optim} function in \texttt{R} with the \texttt{"L-BFGS-B"} algorithm. The initial values are $\rho=\phi_1=0$ and $\sigma^2=1$.
    \item[Step 2. ] Use $(\hat{\rho}, \hat{\phi}_1, \hat{\sigma}^2)$ to compute
    \[
    \hat{\pmb{\beta}} = 
    \Big( \mathbf{Z}^\top \mathbf{A}^{-1} \mathbf{C}^\top \pmb{\Sigma}_{\mathbf{Y}_a}^{-1} \mathbf{C} \mathbf{A}^{-1} \mathbf{Z} \Big)^{-1}
    \mathbf{Z}^\top \mathbf{A}^{-1} \mathbf{C}^\top \pmb{\Sigma}_{\mathbf{Y}_a}^{-1} \mathbf{Y}_a.
    \]
    \item[Step 3. ] Then use $\hat{\pmb{\beta}}, \hat{\rho}, \hat{\phi}_1, \hat{\sigma}^2$ to compute
    \begin{align}
        \widehat{\pmb{Y}} = 
    \widehat{\mathbf{A}}^{-1} \mathbf{Z} \hat{\pmb{\beta}}
    + \widehat{\mathbf{B}}\, \mathbf{C}^\top 
    \big( \mathbf{C} \widehat{\mathbf{B}}\, \mathbf{C}^\top \big)^{-1}
    \Big( \mathbf{Y}_a - \mathbf{C} \widehat{\mathbf{A}}^{-1} \mathbf{Z} \hat{\pmb{\beta}} \Big)\label{ecdesagrega}
    \end{align}
 Equation \eqref{ecdesagrega} corresponds to the best linear unbiased predictor (BLUP). That is the conditional expectation $\mathbb{E}(\mathbf{Y}\vert \mathbf{Y}_a)$ in Equation \eqref{CondNormal}.
By construction, this predictor ensures aggregation coherence since
\[
\mathbf{C} \widehat{\pmb{Y}} 
= \mathbf{C} \tilde{\pmb{Y}} + \mathbf{C} \mathbf{B} \mathbf{C}^\top (\mathbf{C} \mathbf{B} \mathbf{C}^\top)^{-1} (\mathbf{Y}_a - \mathbf{C} \tilde{\pmb{Y}})
= \mathbf{C} \tilde{\pmb{Y}} + (\mathbf{Y}_a - \mathbf{C} \tilde{\pmb{Y}}) = \mathbf{Y}_a.
\]
Therefore, the predicted disaggregated series sums to the observed total. This ensures coherence between the disaggregated estimates and the aggregate data, while minimizing the conditional prediction variance (\cite{chow1971best}, \citet{kelejian2010specification}).
\end{itemize}
We also introduce an alternative with anchorage to Equation \eqref{ecdesagrega} when partial information is available for specific regions and time points. We incorporated known information as linear constraints of the form
\[
\mathbf{H} \pmb{Y} = \mathbf{d},
\]
where $\mathbf{H}$ is a known indicator matrix (with entries in $\{0,1\}$) that selects the positions in $\pmb{Y}$ for which values are known, and $\mathbf{d}$ is the corresponding vector of known values. 
These constraints are incorporated jointly with $\mathbf{C} \pmb{Y} = \mathbf{Y}_a$ by stacking both into an extended constraint system:
\begin{align*}
\widetilde{\mathbf{C}}
\pmb{Y}
=
\begin{bmatrix}
\mathbf{C} \\
\mathbf{H}
\end{bmatrix} 
\pmb{Y}
=
\begin{bmatrix}
\mathbf{Y}_a \\
\mathbf{d}
\end{bmatrix}=\widetilde{\mathbf{Y}}_a .
\end{align*}

We substitute $\mathbf{C}$ and $\mathbf{Y}_a$ in Equation \eqref{ecdesagrega} of Step 3  with their extended versions $\widetilde{\mathbf{C}}$ and $\widetilde{\mathbf{Y}}_a$ respectively,  obtaining 
\begin{equation}
    \widehat{\pmb{Y}} = 
\widehat{\mathbf{A}}^{-1} \mathbf{Z} \hat{\pmb{\beta}} +
\widehat{\mathbf{B}}\, \widetilde{\mathbf{C}}^\top 
\left( \widetilde{\mathbf{C}}\, \widehat{\mathbf{B}}\, \widetilde{\mathbf{C}}^\top \right)^{-1}
\left( \widetilde{\mathbf{Y}}_a - \widetilde{\mathbf{C}}\, \widehat{\mathbf{A}}^{-1} \mathbf{Z} \hat{\pmb{\beta}} \right).\label{ecdesagre22}
\end{equation}
Substituting Equation \eqref{ecdesagrega} for Equation \eqref{ecdesagre22} during Step 3 introduces the anchorage to the model. This anchorage does not require full annual coverage and may include only a few scattered observations across time and regions. 

Now, we show the main results of the proposal.
\begin{theorem}\label{the1}
Under Assumptions \ref{as1}-\ref{as4},
\begin{itemize}
    \item[i)] The parameter vector $\pmb{\theta}=(\pmb{\beta}^\top, \phi_1, \sigma^2, \rho)^\top$ is globally identifiable in Equation \eqref{sart}.
    \item[ii)] The estimator obtained in Equations \eqref{betasar} satisfies that:
    \begin{equation}\label{normalte1}
        \hat{\pmb{\beta}} \xrightarrow[T\to \infty]{d} \mathcal{N}_{k}\left(\pmb{\beta}, \left[\mathbf{Z}^\top(\mathbf{A}^{-1})^{\top}\mathbf{C}^\top\pmb{\Sigma}_{\mathbf{Y}_a}^{-1} \mathbf{C}\mathbf{A}^{-1}\mathbf{Z}\right]^{-1}\right)
    \end{equation}
    \item[iii)]
    \begin{equation}\label{normalte1}
       \widehat{\pmb{Y}}\vert \mathbf{Y}_a\xrightarrow[T\to \infty]{d} \mathcal{N}_{nT}\left(\mathbb{E}(\mathbf{Y}\vert \mathbf{Y}_a), \mathbf{B}- \mathbf{B} \mathbf{C}^\top \pmb{\Sigma}_{\mathbf{Y}_a} ^{-1} \mathbf{C} \mathbf{B}+\mathbf{M}Var(\pmb{\beta})\mathbf{M}^\top\right)
    \end{equation}
    where 
    $\mathbf{M}=\mathbf{A}^{-1}\mathbf{Z}-\mathbf{B}\mathbf{C}^\top \pmb{\Sigma}_{\mathbf{Y}_a}^{-1}\mathbf{CA}^{-1}\mathbf{Z}$.
\end{itemize}
\end{theorem}
\begin{proof}
    See \ref{apenA}
\end{proof}

\subsection{Comparison between the Proposed and Naive Estimators}\label{secYnaive}
In this section, we compare the proposed disaggregation estimator with the naive estimator based on block averages. 
The goal is to identify the conditions under which both estimators yield similar results and, consequently, when the spatially corrected version provides a significant improvement. 

Let the naive predictor of $\mathbf{Y}$ be defined as
\begin{align}
    \hat{\mathbf{Y}}^{1} = \frac{1}{n}\mathbf{C}^\top\mathbf{Y}_a \label{naive}
\end{align}
This estimator corresponds with the temporal vector of spatial means, obtained by averaging observations within each time period.

Now, we present the results that explain when our model outperforms the naive estimator.
\begin{theorem}\label{the2}
    Under Assumptions \ref{as1}-\ref{as4}, let
    \begin{align*}
        \hat{\mathbf{Y}}^{1} &= \frac{1}{n}\mathbf{C}^\top\mathbf{Y}_a,\qquad\mathrm{MSE}(\hat{\mathbf{Y}}^{1},\mathbf{Y})= \frac{1}{nT}\sum_{i=1}^{nT}  (\hat{Y}^{(1)}_{i}-Y_{i})^2,\\
        \Delta &= \mathbb{E}(\mathrm{MSE}(\hat{\mathbf{Y}}^{1}, \mathbf{Y})\mid \mathbf{Y}_a) -\mathbb{E}(\mathrm{MSE}(\hat{\mathbf{Y}}, \mathbf{Y})\mid \mathbf{Y}_a), 
    \end{align*}
    then the following holds
    \begin{align}
     \mathrm{Bias}(\hat{\mathbf{Y}}\mid \mathbf{Y}_a) &=\mathbf{0},\nonumber\\
         \mathrm{Bias}(\hat{\mathbf{Y}}^{1}\mid \mathbf{Y}_a) &=-(\mathbb{I}_{nT}-\mathbf{P})\pmb{\mu_{1a}},\label{muses1}\\
         \mathbb{E}(\mathrm{MSE}(\hat{\mathbf{Y}}^{1},\mathbf{Y})\mid \mathbf{Y}_a)&= \|(\mathbb{I}_{nT}-\mathbf{P})\pmb{\mu}_{1a}\|^{2}
+ \operatorname{tr}\big((\mathbb{I}_{nT}-\mathbf{P})(\mathbb{I}_{nT}-\mathbf{P}_{1})\mathbf{B}\big),\nonumber\\
\mathbb{E}(\mathrm{MSE}(\hat{\mathbf{Y}},\mathbf{Y})\mid \mathbf{Y}_a) &= \operatorname{tr}((\mathbb{I}_{nT}-\mathbf{P}_{1})\mathbf{B}) +\mbox{tr}(\mathbf{M}\mbox{Var}(\hat{\pmb{\beta}})\mathbf{M}^\top),\nonumber\\
 0\leq \Delta &=\pmb{\mu}^\top(\mathbb{I}_{nT}-\mathbf{P})\pmb{\mu}+ \label{deltatate1}\\
 &\; 2k_\rho (\mathbf{Y}_a-\pmb{\mu}_a)^\top (\mathbf{Y}_a-\pmb{\mu}_a)+\label{deltatate2}\\
 &\; \tilde{\pmb{\mu}}^\top \tilde{\pmb{\rho}}-\label{deltatate3}\\
 &\;\mbox{tr}(\mathbf{M}\mbox{Var}(\hat{\pmb{\beta}})\mathbf{M}^\top),\label{deltadeltate4}\\
 \mbox{tr}(\mathbf{M}\mbox{Var}(\hat{\pmb{\beta}})\mathbf{M}^\top)& \leq \frac{k\sigma^2m_\rho\kappa^2_{\mathbf{A}}\kappa^2_{\mathbf{Z}}}{nc_1(1-\vert \phi_1 \vert)^2}, \label{deltatrace}
    \end{align}
where 
\begin{align*}
c_w &= \max_{1\leq j \leq n}\left\{\sum_{i=1}^n w_{ij}\right\}, & \tilde{\pmb{\mu}}&=\begin{pmatrix}
        \sum_{t=1}^T{\mu_{1t}} (Y_{at}-\mu_{at})\\
        \vdots\\
        \sum_{t=1}^T{\mu_{nt}} (Y_{at}-\mu_{at})
    \end{pmatrix}, \\
    \mathbf{P}& = \mathbb{I}_{T} \otimes \frac{1}{n}\mathbf{J}_{n}, &
    \mathbf{P}_{1}&=\mathbf{BC}^\top \pmb{\Sigma}_{\mathbf{Y}_a}^{-1}\mathbf{C},\\
    \mathbf{F}&=\mathbb{I}_n-\rho\mathbf{W}, & m_\rho &=\mathbf{1}_n^\top \mathbf{F}^{-1}(\mathbf{F}^{-1})^\top \mathbf{1}_n,\\
      c_1&=\min_{\mathbf{0}\neq\mathbf{y}\in \langle col(\mathbf{A}^{-1}\mathbf{Z})\rangle} \frac{\|\mathbf{P}\mathbf{y}\|^2}{\|\mathbf{y}\|^2} , &
    \kappa^2_{\mathbf{A}}&=\frac{\lambda_{\max}\left(\left(\mathbf{AA}^\top\right)^{-1}\right)}{\lambda_{\min}\left(\mathbf{\left(\mathbf{AA}^\top\right)}^{-1}\right)},\\
    \kappa^2_{\mathbf{Z}}&=\frac{\lambda_{\max}\left(\mathbf{Z}^\top\mathbf{Z}\right)}{\lambda_{\min}\left(\mathbf{Z}^\top\mathbf{Z}\right)}, &
    \mathbf{D}_\rho&=\mathbf{F}^{-1}(\mathbf{F}^{-1})^\top m_\rho^{-1}\mathbf{1}_n,\\
    k_\rho&= \mathbf{D}_\rho^\top\left(\mathbb{I}_n -\frac{1}{n} \mathbf{J}_n\right)\mathbf{D}_\rho, & \tilde{\pmb{\rho}}&=\left(\mathbf{D}_\rho-\frac{1}{n}\mathbf{1}_n\right).
\end{align*}
\end{theorem}
\begin{proof}
    See \ref{AppenB}
\end{proof}
Concretely, $c_w=\|\mathbf{W}\|_1$ is the maximum column sum of $\mathbf{W}$; $\tilde{\pmb{\mu}}$ denotes the spatially aggregated mean vector, obtained by averaging across spatial units while accounting for temporal differences; $\mathbf{P}$ is the projection matrix over temporal means; $\mathbf{P}_1$ is the projection over the temporal aggregated variance $\pmb{\Sigma}_{Y}$; $m_\rho$ is the effect of spatial variance over the temporal aggregated variance; $\lambda_{\min}(\cdot)$ and $\lambda_{\max}(\cdot)$ are the minimum and maximum eigenvalue function, respectively. 

In Equation~\eqref{deltatate1}, 
$\pmb{\mu}^\top(\mathbb{I}_{nT}-\mathbf{P})\pmb{\mu}$, 
is always nonnegative, and measures the spatial heterogeneity of the mean component 
$\pmb{\mu}$. 
Since $\mathbf{P}$ replaces each observation by its spatial average within the corresponding period, the term $(\mathbb{I}_{nT}-\mathbf{P})\mathbf{Z}$ represents the demeaned component of $\mathbf{Z}$. Hence, $\mathbf{Z}^\top(\mathbb{I}_{nT}-\mathbf{P})\mathbf{Z}$ quantifies the amount of spatial heterogeneity present in the explanatory variables $\mathbf{Z}$ once the temporal block means have been removed.
Consequently, the term $\pmb{\mu}^\top(\mathbb{I}_{nT}-\mathbf{P})\pmb{\mu}$ is large when $\mathbf{Z}$ exhibits strong spatial variation across regions within a given period, and small when $\mathbf{Z}$ has similar values across space. In particular, $\pmb{\mu}^\top(\mathbb{I}_{nT}-\mathbf{P})\pmb{\mu}$ is small if $\|\pmb{\beta}\|$ is small or the regressors $\mathbf{Z}$ yield spatially homogeneous values.

Since $\pmb{\mu} = \mathbf{A}^{-1}\mathbf{Z}\pmb{\beta}$, the spatial filter $\mathbf{A}^{-1}$ alters this behavior by transmitting local shocks across neighboring regions according to the spatial dependence parameter $\rho$. As $|\rho|$ increases, $\mathbf{A}^{-1}$ amplifies spatial spillovers, leading to stronger inter-regional dependence and a larger value of $\pmb{\mu}^\top(\mathbb{I}_{nT}-\mathbf{P})\pmb{\mu}$. In contrast, when $\rho$ approaches zero, $\mathbf{A}^{-1}$ tends to the identity matrix and the spatial heterogeneity of $\pmb{\mu}$ mirrors that of $\mathbf{Z}$ itself.

The vector $\mathbf{D}_\rho=\mathbf{F}^{-1}(\mathbf{F}^{-1})^\top m_\rho^{-1}\mathbf{1}_n$ depends on the spatial autoregressive parameter $\rho$ through $\mathbf{F}=\mathbb{I}_n-\rho\mathbf{W}$. 
When $\rho=0$, $\mathbf{F}=\mathbb{I}_n$ and $\mathbf{D}_\rho = m_\rho^{-1}\mathbf{1}_n=\frac{1}{n}\mathbf{1}_n$, that is, $\mathbf{D}_\rho$ is proportional to the unit vector. In this case, there is no spatial propagation, and all regions contribute equally to the aggregated mean. As $|\rho|$ increases, $(\mathbb{I}_n-\rho\mathbf{W})^{-1}$ introduces spatial spillovers across neighboring units, so the elements of $\mathbf{D}_\rho$ deviate from uniformity, reflecting heterogeneous spatial influence. Therefore, $\mathbf{D}_\rho$ is approximately proportional to $\mathbf{1}_n$ 
when $\rho$ is close to zero and becomes increasingly heterogeneous as spatial dependence strengthens. The value of $k_\rho$ in Equation \eqref{deltatate2} is close to $0$ if $\rho$ is close to $0$.

The value of $(\mathbf{Y}_a-\pmb{\mu}_a)^\top (\mathbf{Y}_a-\pmb{\mu}_a)$ in Equation \eqref{deltatate2} is close to $0$ if the aggregated series are close to their means ($\mathbf{Y}_a\approx \pmb{\mu}_a$). This occurs when $\vert\phi_1 \vert$ is close to 1 and $\sigma^2$ is small. 

The term $\tilde{\pmb{\mu}}^\top \tilde{\pmb{\rho}}$ of Equation \eqref{deltatate3} has expected value $0$ because $\mathbb{E}(\mathbf{Y}_a-\pmb{\mu}_a)=0$. 
If ${\rho}=0$, we have $\mathbf{D}_\rho=\frac{1}{n}\mathbf{1}_n$ which yields $\tilde{\pmb{\mu}}^\top \tilde{\pmb{\rho}}=0$. If the value $\tilde{\pmb{\mu}}^\top \tilde{\pmb{\rho}}$ is positive, the estimator proposed in this work is better than the naive one. If the value $\tilde{\pmb{\mu}}^\top \tilde{\pmb{\rho}}$ is negative,
we can estimate it using Theorem \ref{the1}.

The upper bound of the trace increases as $|\rho|$ approaches $1$, and as $|\phi_1|$ tends to $1$, which leads to an increase of the variance associated with the estimation of $\pmb{\beta}$. When the number of parameters $k$ is large relative to $n$, or when $\sigma^2$ is large, the bound also increases. Nevertheless, this bound remains estimable under the conditions established in Theorem~\ref{the1}.

In summary, the proposed disaggregation series approximates to the naive estimator when the following conditions hold simultaneously:
\begin{itemize}
\item The regression coefficients are small, i.e., $\pmb{\beta}^\top\pmb{\beta}$ is close to $0$.
\item The covariates are spatially homogeneous across regions and periods, that is, $\|\mathbf{Z}-\tfrac{1}{n}\mathbf{C}^\top\mathbf{C}\mathbf{Z}\|\approx 0$.
\item The spatial dependence parameter $\rho$ is close to $0$.
\item The temporal persistence is high ($|\phi_1|\approx 1$) and the innovation variance $\sigma^2$ is small.
\item The interaction term $\tilde{\pmb{\mu}}^\top \tilde{\pmb{\rho}}$ in Equation \eqref{deltatate3} is negative and far from $0$.
\end{itemize}

When these conditions do not hold jointly, the proposed disaggregation method outperforms the naive estimator ($\Delta>0$), as it improves the capture of spatial and temporal dynamics. Each of these conditions can be analytically derived from the model structure and empirically verified using the observed covariates $\mathbf{Z}$ and estimated parameters. However, satisfying all five conditions simultaneously would imply an almost static system with minimal spatial heterogeneity and temporal variability—an unrealistic situation in most applied contexts. Therefore, it is very unlikely that the naive estimator performs as well as the proposed one in practical scenarios. Moreover, since the proposed model consistently dominates the naive specification, its performance is, in the worst case, lower-bounded by that of the naive estimator.

\section{Simulation}

To empirically validate the quality of the model, we perform a simulation with synthetic data in R.

\subsection{Synthetic data generation}

We generate synthetic data with parameters $n,T,\rho,\phi_1,\theta,\beta_0,\beta_1,\sigma$ in the following way:

To start, we fix the number of municipalities $n$ as a square number. We position spatially the municipalities in a grid as presented in Figure~\ref{fig:sim_1}. We construct the $n\times n$ adjacency matrix $\mathbf{W}$ as $w_{ij}=1$ when municipalities $i$, and $j$ are adjacent, and $0$ otherwise.

\begin{figure}[ht]
	\centering
	\begin{tikzpicture}[scale=1]
		\foreach \row in {0,...,3} {
			\foreach \col in {0,...,3} {
				\pgfmathtruncatemacro{\n}{(3-\row)*4 + \col + 1}
				
				\draw (\col, \row) rectangle ++(1,1);
				
				\node at (\col+0.5,\row+0.5) {\small\bfseries\n};
			}
		}	
	\end{tikzpicture}
	\caption{Example of spatial positioning of synthetic data with $4^2=16$ municipalities. Municipality $6$ is adjacent to $1,2,3,5,7,9,10,11.$}
	\label{fig:sim_1}
\end{figure}

We continue with the $nT\times 2$ matrix $\mathbf{Z}$, where first column entries are one and each entry $z_{i2}$ is the realization of a random variable $ X \sim \mathcal{U}[0,1] $.

The generation of $AR(1)$ 
data is performed with the \verb|arima.sim| instruction, part of the \verb|stats| package, version 4.4.3.

We obtain the synthetic data $\mathbf{Y}$ as the result of 
\[\mathbf{Y}=\mathbf{A}^{-1}\mathbf{Z}\pmb{\beta}+\mathbf{A}^{-1}\mathbf{U},\]
where $\pmb{\beta} = [\beta_0,\beta_1]^T$ and 
\[\mathbf{A}^{-1}=\mathbb{I}_T\otimes (\mathbb{I}_n-\rho \mathbf{W})^{-1}.\]

The usage of $\mathbf{Z,Y,W,U}$ corresponds with the notation of Equation~\eqref{align:model_data}. To fasten the computation the package \verb|furrr| was used to parallelize the process in a local Linux Server.

\subsection{Simulation parameters and metrics}

For all combinations of values:
\begin{multicols}{2}
\begin{itemize}
	\item $n\in \{9,16,25,36,49,64\}$ 
	\item $T\in \{12,24,36,\dots,144\}$
	\item $\rho\in \{0,\pm0.25,\pm0.5,\pm0.75\}$
	\item $\phi_1\in \{0,\pm0.25,\pm0.5,\pm0.75\}$
	\item $\beta_0=1$
	\item $\beta_1=\{0,0.5,1,5,10,50,100\}$
	\item $\sigma=\{0.1,\sqrt{0.1},1\}$
\end{itemize}
\end{multicols}

we have generated synthetic data and executed the proposed model to see the quality of our results. 
We have considered the following metrics
\begin{align}
  R^2 &= 1 - \frac{ \sum_{i=1}^{n} \sum_{t=1}^{T} (Y_{it} - \hat{Y}_{it})^2 }
                  { \sum_{i=1}^{n} \sum_{t=1}^{T} (Y_{it} - \bar{Y})^2 },\nonumber \\
  MAPE &= \frac{1}{nT} \sum_{i=1}^{n} \sum_{t=1}^{T}
         \left| \frac{Y_{it} - \hat{Y}_{it}}{Y_{it}} \right| \cdot 100,\nonumber \\
  RRMSE &= \frac{RMSE}{\bar{Y}}
        = \frac{1}{\bar{Y}} \sqrt{ \frac{1}{nT} \sum_{i=1}^{n} \sum_{t=1}^{T} (Y_{it} - \hat{Y}_{it})^2 }, \nonumber\\
  RMSE &= \sqrt{ \frac{1}{nT} \sum_{i=1}^{n} \sum_{t=1}^{T} (Y_{it} - \hat{Y}_{it})^2 }, \nonumber\\
 \chi^2 &= \sum_{i=1}^n\left(\frac{\sum_{t=1}^T(Y_{it}-\hat{Y}_{it})}{\sum_{t=1}^T Y_{it}}\right).\label{distChi}\end{align}

Due to the high number of parameter combinations ($6\cdot 12\cdot 7\cdot 7\cdot 7\cdot 3=74088$) we grouped the results in $3$ categories depending on the value of $\frac{\beta_1}{\sigma^2}$ as by the signal-to-noise ratio. Lower values indicate weaker covariate effects relative to the residual variance. The specific choice of values was made to balance the number of results per category (see Table~\ref{tab:category_values}).
\[
\text{ratio} = \frac{\beta_1}{\sigma^2}
\quad 
\text{with}
\quad 
\begin{cases}
\text{Low}         & \text{if } \text{ratio} < 5, \\
\text{Medium}      & \text{if } 5 \leq \text{ratio} < 50, \\
\text{High}        & \text{if } 50 \leq \text{ratio} \leq 500, \\
\text{Very High}   & \text{if } \text{ratio} > 500.
\end{cases}
\]

\begin{table}[ht]
	\centering
		\caption{Number of results per category.}
	\begin{tabular}{lc}
		\hline
		{Category} & {Count} \\
		\hline
		Low        & 21168 \\
		Medium     & 17640 \\
		High       & 21168 \\
		Very High  & 14112 \\
		\hline
	\end{tabular}
	\label{tab:category_values}
\end{table}

\begin{figure}[ht]
	\centering
	\includegraphics[width=14cm]{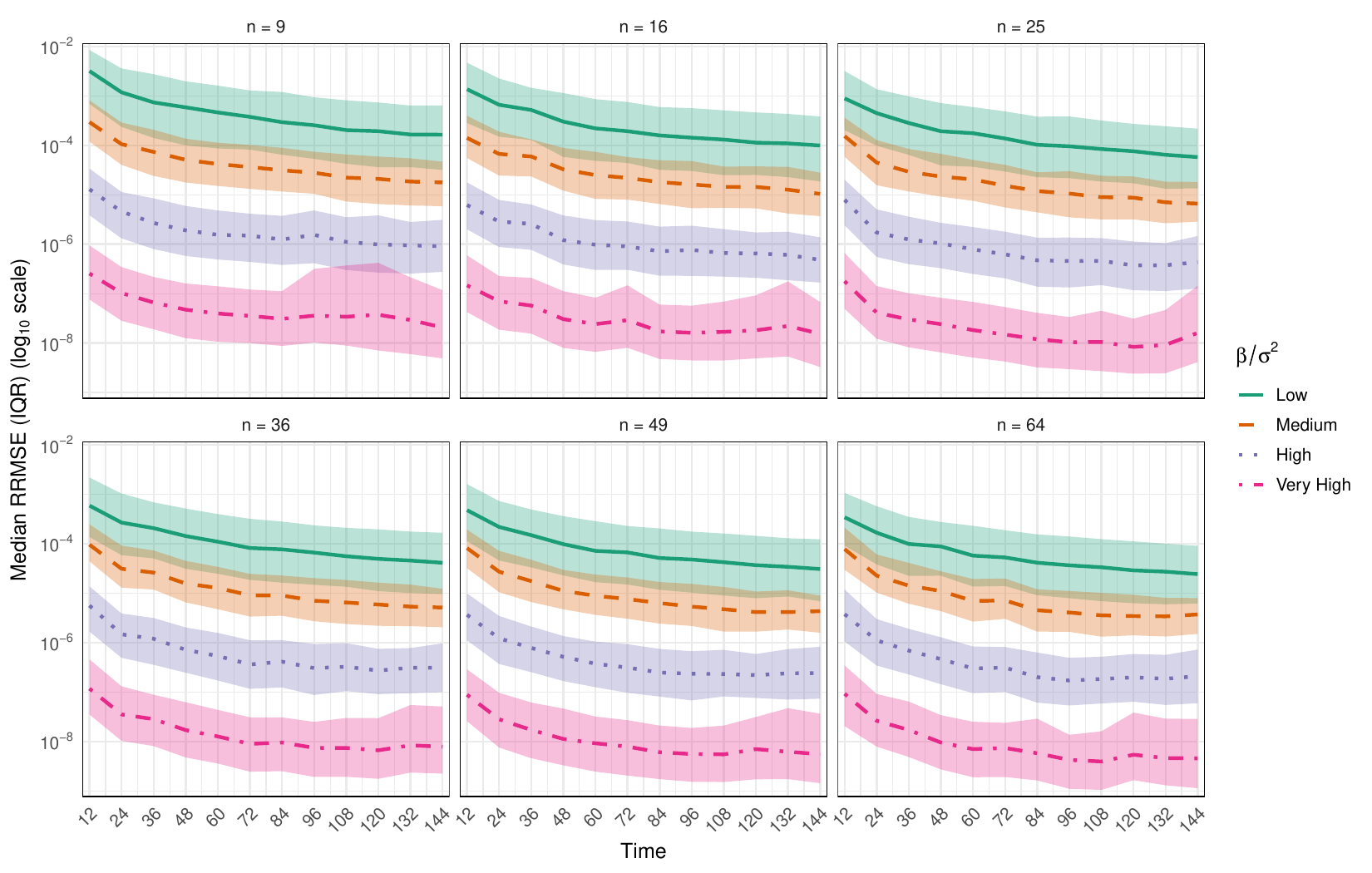}  
	\caption{Median relative root mean squared error (RRMSE) with interquartile ranges (shaded ribbons) plotted on a log scale, across different time lengths $Time$ and sample sizes $n$. Colors represent different signal-to-noise ratio levels measured by $\beta/\sigma^2$.}
	\label{fig:ntRMSE}
\end{figure}

Figure \ref{fig:ntRMSE} shows that the behavior of the median RRMSE and the quartiles is similar regardless of the size of the spatial grid according to the quotient between $\frac{\beta_1}{\sigma^2}$. It is clear that the higher this value, the better the reconstruction of the time series. It is also shown that as $T$ grows, there is a slight decrease in the RRMSE of the disaggregation, although it is a smooth logarithmic trend.
\begin{table}[ht]
\centering
{\footnotesize
\caption{Average performance metrics by covariate strength level and number of spatial units ($n$).}
\label{tab:metrics}
\begin{tabular}[t]{lrrrrrr}
\toprule
$\beta_1/\sigma^2$ & n & RMSE & $\log$ RRMSE & MAPE & $R^2$ & $\chi^2$\\
\midrule
\multirow{6}{*}{Low}
  & 9 & 0.856 & -8.030 & 5.124 & -0.036 & 1.515\\
 & 16 & 0.834 & -8.591 & 3.377 & -0.002 & 1.851\\
 & 25 & 0.845 & -9.011 & 4.590 & -0.031 & 1.560\\
 & 36 & 0.843 & -9.384 & 6.488 & -0.047 & 1.553\\
 & 49 & 0.843 & -9.693 & 4.384 & -0.048 & 1.539\\
  & 64 & 0.859 & -9.941 & 6.599 & -0.110 & 1.629\\
   \hline
\multirow{6}{*}{Medium}
  & 9 & 0.716 & -10.162 & 1.143 & 0.704 & 0.371\\
 & 16 & 0.754 & -10.626 & 0.894 & 0.658 & 0.446\\
& 25 & 0.786 & -11.018 & 0.897 & 0.612 & 0.415\\
& 36 & 0.819 & -11.344 & 0.950 & 0.573 & 0.429\\
& 49 & 0.826 & -11.638 & 1.076 & 0.567 & 0.428\\
& 64 & 0.895 & -11.833 & 1.382 & 0.450 & 0.474\\
      \hline
\multirow{6}{*}{High}
 & 9 & 1.177 & -13.214 & 0.285 & 0.956 & 0.320\\
 & 16 & 1.080 & -13.739 & 0.230 & 0.953 & 0.163\\
 & 25 & 1.118 & -14.096 & 0.230 & 0.950 & 0.256\\
 & 36 & 1.077 & -14.420 & 0.360 & 0.942 & 0.142\\
 & 49 & 1.172 & -14.704 & 0.191 & 0.938 & 0.129\\
 & 64 & 1.502 & -14.864 & 0.207 & 0.906 & 0.277\\
      \hline
\multirow{6}{*}{Very High}
   & 9 & 1.513 & -16.493 & 0.120 & 0.969 & 0.627\\
  & 16 & 1.365 & -17.143 & 0.069 & 0.961 & 0.311\\
   & 25 & 1.013 & -17.590 & 0.076 & 0.968 & 0.323\\
   & 36 & 0.671 & -17.925 & 0.085 & 0.983 & 0.092\\
  & 49 & 0.866 & -18.197 & 0.139 & 0.974 & 0.100\\
 & 64 & 1.495 & -18.310 & 0.071 & 0.951 & 0.344\\
\bottomrule
\end{tabular}
 }
\end{table}

The small values show that the disaggregation yields very low metric results. Table \ref{tab:metrics} shows the summary metrics for each of the ratio levels between $\frac{\beta_1}{\sigma^2}$. Since the RMSE is a non-relative measure, it increases as the ratio increases. This is explained because as the ratio increases, the variable $Y_{it}$ in the simulation takes on larger values. Therefore, by relativizing this RMSE, the RRMSE and the $\log RRMSE$ are shown to illustrate the difference between the models. This is much smaller as the ratio increases, as seen in Figure~\ref{fig:ntRMSE}. The MAPE is also shown, where the reduction is even more significant, with values less than 1\% for the High and Very High values. The $R^2$ is also shown, which obtains very good values for High and Very High. It presents medium-high $R^2$ values when the ratio is Medium, and negative values close to zero when the ratio is Low. This is explained because, when the variables $\pmb{Z}$ are not informative in the proposed model, the disaggregation is similar to performing a disaggregation with the mean, that is, a proxy for naivety (See $\hat{\mathbf{Y}}^{1}$ estimator in Section \ref{secYnaive}).

The value of $\chi^2$ proposed on the Equation \eqref{distChi} weights the goodness of fit for each spatial unit. This is of particular interest because it allows us to assess whether the spatial units are well disaggregated. It is shown that the model disaggregates all variables very well for each spatial unit, even with large $n$ sizes.

\begin{figure}[ht]
	\centering
\includegraphics[width=14cm]{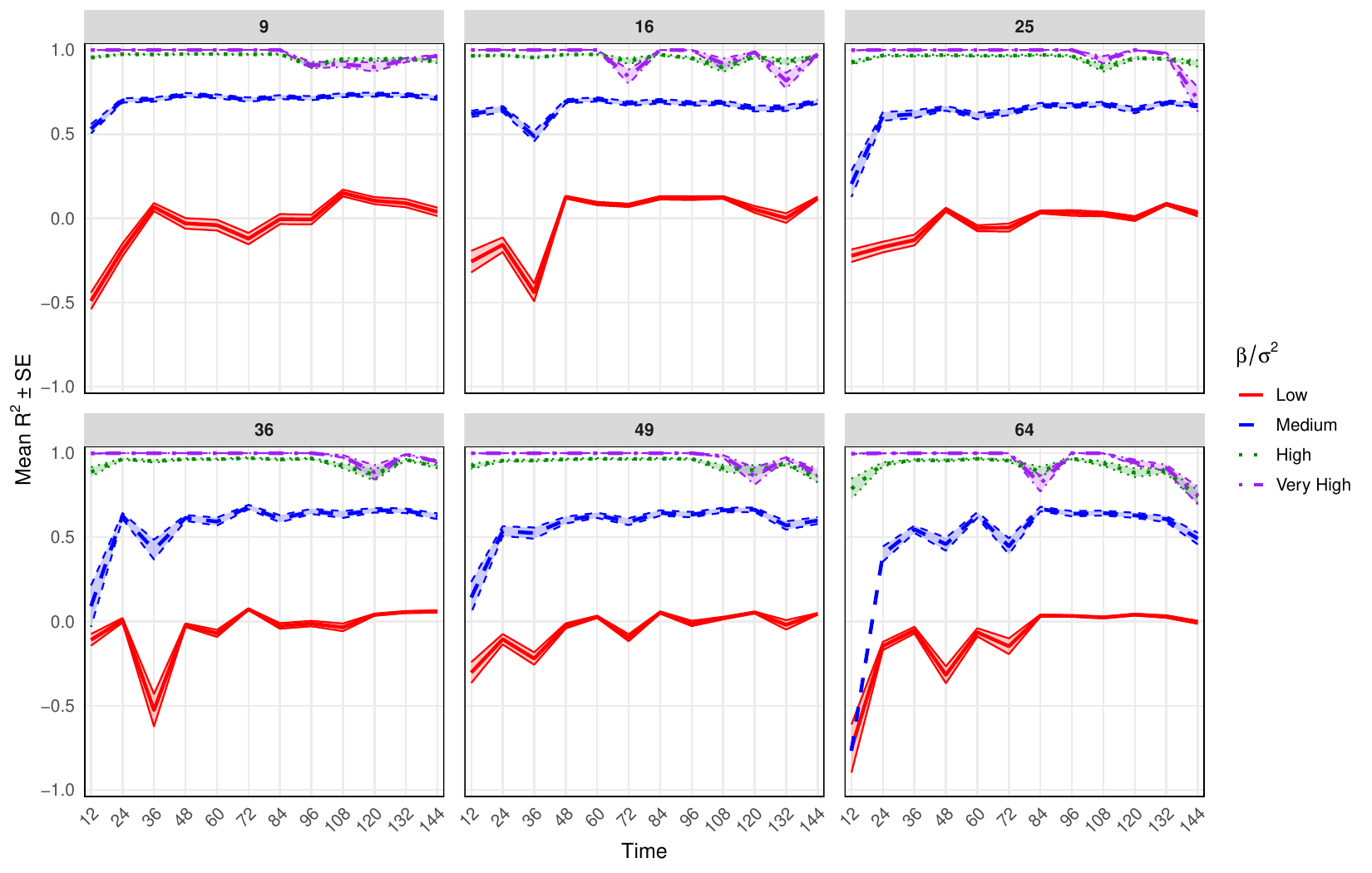} 

	\caption{Mean observed $R^2$ with standard error bands across different time lengths (Time), grouped by the number of spatial units ($n$) shown in facets. 
Colours represent levels of the ratio $\beta/\sigma^2$. }
	\label{fig:ntR2}
\end{figure}

What was stated above regarding $R^2$ is reaffirmed in Figure \ref{fig:ntR2}. In Table \ref{tab:metrics}, it is shown that the behavior of this metric is very similar regardless of the value of $n$.
Furthermore, only in the case of a low coefficient $\frac{\beta_1}{\sigma^2}$ does the model yield mediocre results, because the information provided by the covariates is too limited to reconstruct the series.

It should be noted that all the metrics shown above were calculated using the average of all $\rho$ values. This demonstrates the consistency of the method regardless of the value of the spatial correlation.
However, we will now discuss the behavior of the RRMSE with respect to $\rho$, the spatial component of the model.
\begin{figure}[ht]
	\centering
\includegraphics[width=14cm]{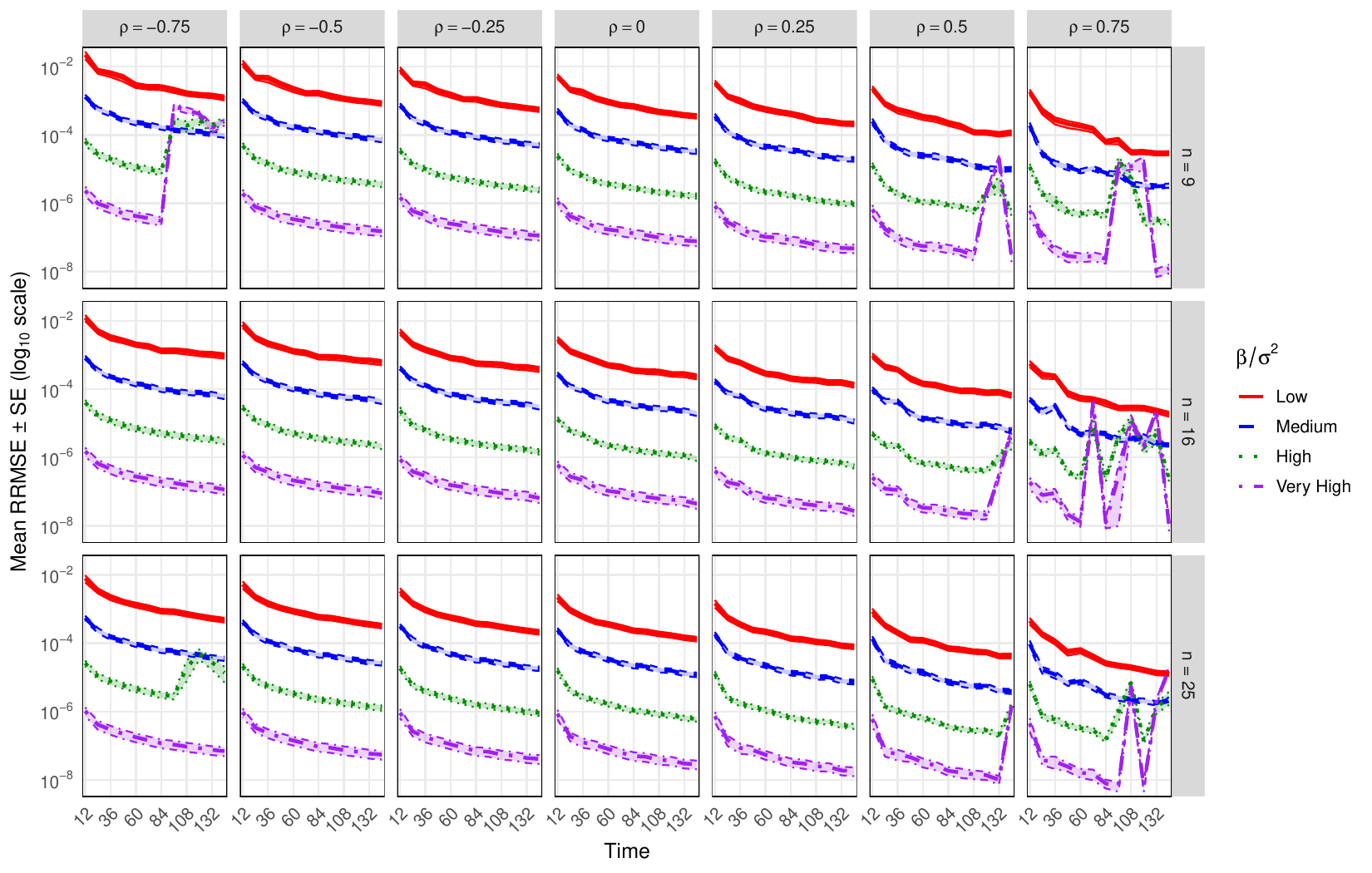}

	\caption{Mean relative root mean squared error (RRMSE) with standard error bands on a log scale across different time lengths ($Time$), grouped by the number of samples ($n=\{9, 16, 25\}$) and spatial correlation ($\rho$) shown in facets. Colors represent levels of the ratio $\beta/\sigma^2$. }
	\label{fig:ntrhoRMSE}
\end{figure}

Figure \ref{fig:ntrhoRMSE} shows the mean RRMSE values on a logarithmic scale. They are subdivided by the number of spatial units. Here it is shown for 9, 16, and 25. It is observed that the behavior according to the quotient $\beta/\sigma^2$ is similar to what we had already observed in the previous figures; that is, the higher the quotient, the better the breakdown of the time series. But this behavior only occurs for $\rho$ values of -0.5, 0.25, 0, and 0.25. When $\rho$ is $\pm 0.75$, very anomalous behavior occurs when $n$ is 9. And when $\rho$ is 0.5, anomalous behavior occurs with very large $T$ values. As for the value of $\rho=0.75$, when the $T$ is greater than 36, the model is completely damaged at high ratios. This is justified by the results obtained in Theorem \ref{the2}.

\begin{figure}[ht]
	\centering
	\includegraphics[width=14cm]{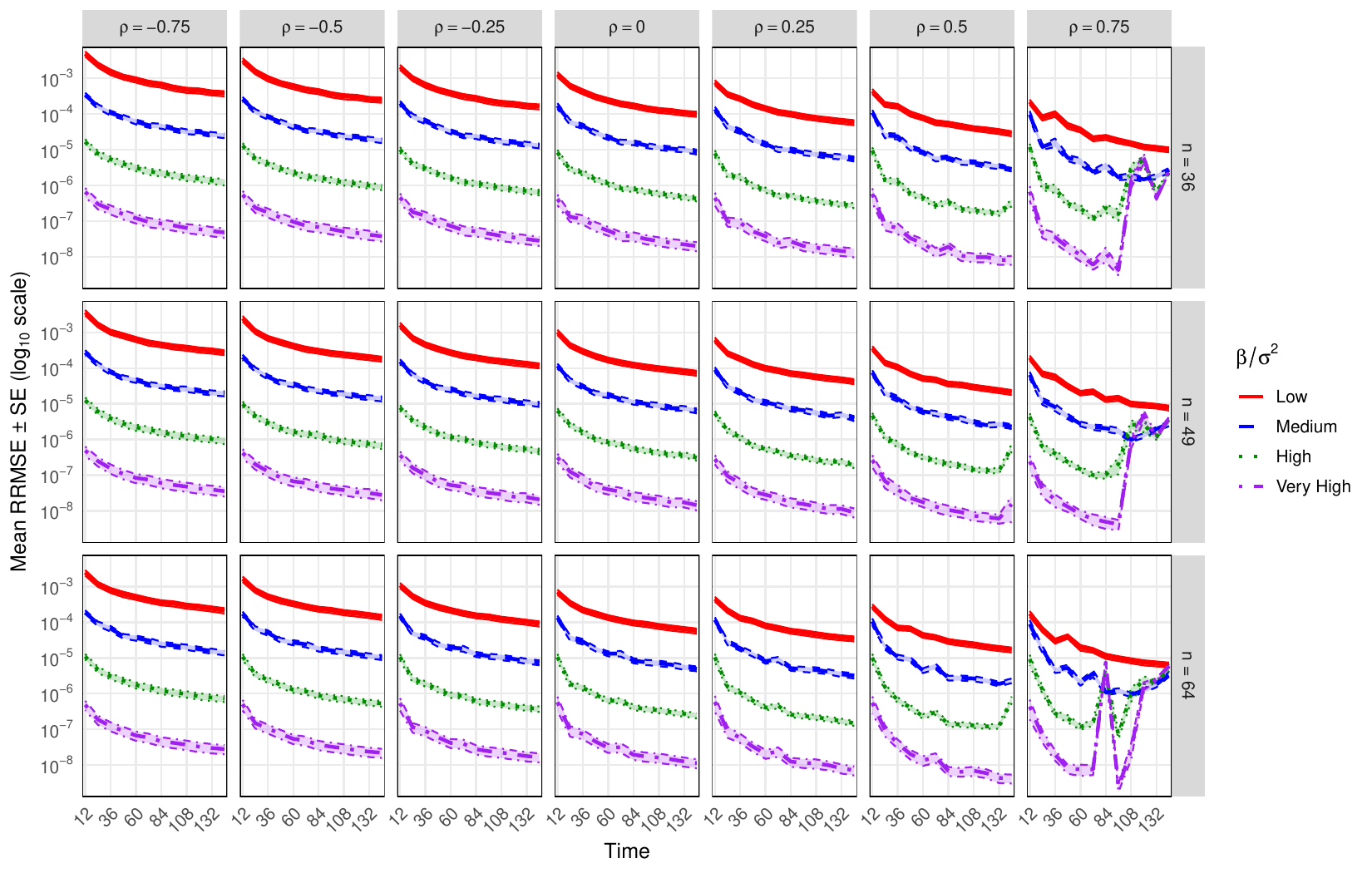}
	\caption{Mean relative root mean squared error (RRMSE) with standard error bands on a log scale across different time lengths (Time), grouped by the number of samples ($n=\{36, 49, 64\}$) and spatial correlation ($\rho$) shown in facets. Colors represent levels of the ratio $\beta/\sigma^2$.}
	\label{fig:ntrhoRMSE1}
\end{figure}
In Figure \ref{fig:ntrhoRMSE1} it is shown analogously to Figure \ref{fig:ntrhoRMSE} for 36, 49, and 64. It is observed that the behavior is good for all values of $\rho$ except when this value is 0.75 and $T$ is greater than 48. With these last two figures, it is observed that the spatial part alters the model when it is very strong, because the matrix $\mathbf{A}$ is close to being singular, and therefore, presents estimability problems. Furthermore, in a practical sense, if the variables $\pmb{Z}$ do not capture the spatiality, they will make the model unstable. Therefore, this model should be applied with caution if the estimator of $\rho$ is close to 1 and the sample size in the time series is medium or large. In all other scenarios, the disaggregation is as good as the covariates that allow the disaggregation of the series of interest.
\section{Application- GDP in Spain}

This document describes the auxiliary variables employed to disaggregate the Gross Domestic Product (GDP) in Spain at the level of Autonomous Communities (NUTS-2) from 2002 to 2023. These covariates are selected for their economic relevance, availability at the regional level, and statistical relationship with the GDP components. All data sources are official and publicly available.

\begin{itemize}
\item \textbf{Industrial Production Index (IPI).} The IPI is a monthly short-term indicator measuring changes in the physical volume of industrial production in Spain (excluding construction), covering sections B (Mining and quarrying), C (Manufacturing), D (Electricity and gas supply), and division 36 of section E (Water supply; sewerage, waste management, and remediation activities) of CNAE-2009. Based on a survey of over 11 500 establishments, it strips out price effects to track real output dynamics \cite{INE_IPI2025}.

\item \textbf{Population.} The annual Municipal Register provides resident counts by Autonomous Community; population size scales production and consumption, thereby influencing regional GDP levels \cite{INE_Pop2025}.

\item \textbf{Industrial Energy Consumption.} The ENC survey collects annual data on electricity, natural gas, fuel oil, renewables, petroleum products, and other fuels used by industrial firms with at least 20 employees in CNAE-2009 sections B and C. It serves as a proxy for industrial intensity \cite{INE_ENC2022}.

\item \textbf{Business Demography.} Monthly statistics on company births, dissolutions, and capital changes yield new firm counts, subscribed capital, and disbursed capital—key indicators of business dynamism and regional investment activity \cite{INE_BusDemo2022}.

\item \textbf{Household Food Expenditure.} Annual monetary expenditure on food by households, disaggregated by Autonomous Community, captures regional consumption patterns relevant for GDP via the expenditure approach~\cite{MAPA_Food2022}.

\item \textbf{Labour Force and Activity Rates.} Quarterly Labour Force Survey data on the economically active population by sex and activity rate reflect the availability and participation of human capital across regions \cite{INE_LFS2022}.
\end{itemize}

\vspace{0.5cm}
These variables serve as proxies of sectoral, demographic, and labour market conditions across regions and over time, enabling a robust estimation of the regional GDP trajectories within the disaggregation framework.

\subsection{Spatial Weight Matrices}

Instead of relying on conventional definitions of spatial proximity—such as contiguity or fixed-distance thresholds—we developed a custom spatial weight matrix designed to reflect spatial relationships based on the socio-economic and demographic structure of each region. To construct this matrix, we employed Gower distance, which is particularly well-suited for spatial analysis involving variables of mixed types. Unlike Euclidean distance, Gower distance accommodates heterogeneous data structures, making it especially appropriate for comparing regions with complex socio-economic profiles. For example, \cite{comber2011spatial} applied Gower distance to analyze spatial disparities in access to health services by incorporating multiple socio-economic dimensions, while \cite{dai2011geographic} used it to assess inequalities in food accessibility based on a combination of demographic and geographic indicators. Originally introduced by \cite{gower1971general}, this distance metric offers a flexible and interpretable framework for capturing spatial similarity in contexts where conventional proximity measures may be inadequate.

\begin{figure}[ht]
	\centering
	\includegraphics[width=12cm]{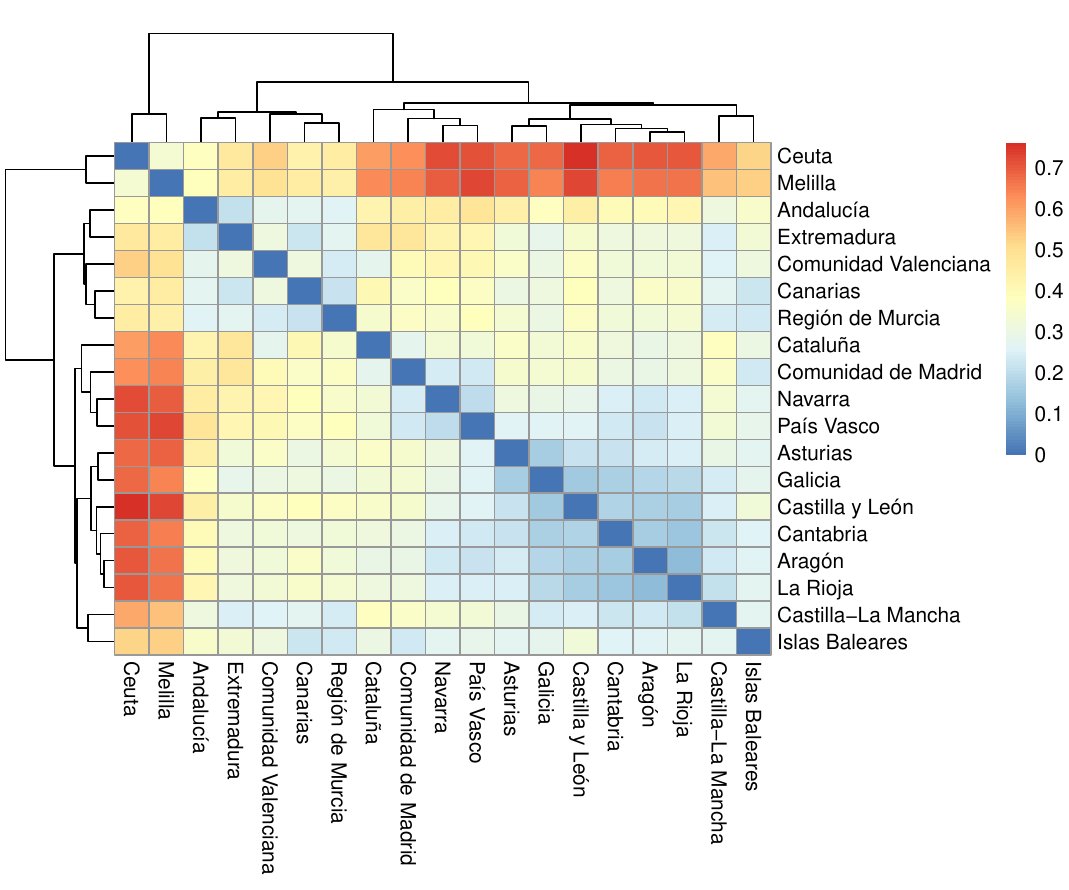}
	\caption{Heatmap of Gower distances between Spanish autonomous communities. Each cell represents the dissimilarity between a pair of regions based on eleven socio-economic and demographic variables for the year 2021. Darker blue shades indicate higher similarity (lower distance), while lighter red shades reflect greater dissimilarity (higher distance). The dendrogram shows the hierarchical clustering of regions with similar profiles.}
	\label{fig:Gower}
\end{figure}

To compute the Gower distance between each pair of autonomous communities, we used a set of eleven indicators corresponding to the year 2021: population density, birth rate per 1,000 inhabitants, mortality rate per 1,000 inhabitants, life expectancy, risk of poverty, Human Development Index (HDI), Gini index, total public debt (in millions of euros), public deficit (in millions of euros), unemployment rate according to the Labour Force Survey (LFS), and the Consumer Price Index (CPI). This multidimensional set captures structural differences across regions and supports a more substantive definition of spatial similarity. The distance matrix was calculated using the \texttt{daisy()} function from the \texttt{cluster} package in R \cite{cluster2024}, which offers a robust implementation for mixed data types. All variables were assigned equal weights, ensuring that each contributed equally to the overall dissimilarity measure.
In Figure~\ref{fig:Gower}, we present the resulting Gower distance matrix between the autonomous communities using a heatmap, where each cell represents the pairwise dissimilarity between two regions.\footnotetext{Note: Autonomous community names are presented in Spanish as per their official designation.} Darker blue tones indicate higher similarity (lower distance), while lighter red tones denote greater dissimilarity (higher distance). A hierarchical clustering dendrogram accompanies the heatmap, grouping regions with similar socio-economic and demographic profiles. This visualization facilitates the identification of clusters of autonomous communities that share comparable structural characteristics.

On the rightmost side of the heatmap, Castilla-La Mancha, Islas Baleares, La Rioja, Aragón, and Cantabria appear closely clustered and share predominantly blue-toned cells, indicating a high degree of similarity in their regional profiles. These regions display comparable values in population density, birth and mortality rates, life expectancy, and unemployment. Similarly, Navarra, the Basque Country, Asturias, Galicia, and Castilla y León are grouped together in the dendrogram, suggesting structural proximity with respect to low poverty risk, high HDI scores, and similar CPI levels.

In contrast, Ceuta and Melilla consistently exhibit reddish shades in their pairwise comparisons with all other autonomous communities, highlighting their distinctiveness. These differences are likely due to their high population densities, elevated birth rates, and markedly different labor market indicators, such as unemployment. Andalucía also shows substantial distances from many other regions, suggesting a differentiated socio-economic profile characterized by higher unemployment, lower HDI, and above-average poverty risk. These patterns underscore the value of incorporating a wide range of indicators when defining spatial similarity, as doing so reveals not only geographic groupings but also deeper structural disparities across regions.

\subsection{Selection of Covariates}
Although some energy-related variables exhibited missing values in certain years and regions, these gaps were relatively limited in both frequency and magnitude. To preserve the internal structure and temporal consistency of each series, imputation was conducted independently for each region and variable using linear interpolation over time. This method, while simple, is widely used when the primary concern is maintaining continuity without introducing artificial trends or external information \cite{LittleRubin2002, Junger2015}. More sophisticated imputation strategies could potentially exploit cross-variable or cross-region correlations, but these were deliberately avoided to prevent leakage of information that could bias the temporal disaggregation model.

Ceuta and Melilla were excluded from the analysis due to their unique administrative and economic profiles, which differ substantially from those of the 17 main Autonomous Communities and may lead to unstable or unrepresentative estimates.

Because many covariates exhibited strong multicollinearity—especially due to the use of multiple correlated economic indicators and energy series—a Principal Component Analysis (PCA) was applied to the standardized covariates. PCA is a classical technique for dimension reduction that transforms a correlated set of variables into a smaller number of uncorrelated components that retain most of the original variance. This transformation improves numerical stability and interpretability in regression-type models, especially when the number of predictors is large relative to the sample size or when predictors are highly collinear \cite{jolliffe2011}. In the context of temporal disaggregation, the use of PCA has been shown to improve estimation by regularizing the predictor space and avoiding overfitting \cite{Durbin1960, proietti2006temporal}.

Instead of selecting covariates based on out-of-sample predictive error—which is not directly available in this context—we relied on structural diagnostics of the fitted models. In particular, we focused on the estimated innovation variance ($\hat{\sigma}^2$) and the autocorrelation parameter ($\hat{\rho}$) from the temporal model. Simulation studies performed in parallel to this application indicated that high values of $\hat{\rho}$ (close to 1) in conjunction with large innovation variances were associated with unstable or biased disaggregation, especially when the series included regions with sparse or weak signal. Consequently, among the different combinations of covariates explored, we prioritized those yielding lower values of $\hat{\sigma}^2$ and moderate values of $\hat{\rho}$, avoiding those approaching the unit root boundary. 

This selection criterion ensures that the model retains a reasonable degree of temporal smoothness without being overly driven by persistence, while also mitigating the propagation of uncertainty during the disaggregation process. The final model included a reduced set of principal components derived from the full pool of covariates, ensuring interpretability and stability without relying on subjective or ad hoc selection mechanisms.

\subsection{Results of disaggregation}

From the full set of available covariates, a subset of six variables was selected for inclusion in the final disaggregation model based on their interpretability and contribution to the explained variance. The selected variables were: Household Food Expenditure, Total Energy Consumption, Number of Active Men, Electricity Consumption, Coal and Derivatives Consumption, and Gas Consumption. This selection captures key aspects of household demand, labor force participation, and energy use, which are all closely related to regional GDP dynamics.

\begin{table}[ht]
\centering
\caption{Principal Component Analysis results: standard deviations, explained variance, and loadings}
\label{tab:PCA_results}
\begin{tabular}{lccccccc}
\toprule
Variable / Statistic           & PC1     & PC2     & PC3     & PC4     & PC5     & PC6     \\
\midrule
\textbf{Standard Deviation}   & 2.196   & 0.842   & 0.563   & 0.320   & 0.196   & 0.103   \\
\textbf{Explained Variance (\%)} & 80.4    & 11.7    & 5.7     & 1.7     & 0.6     & 0.2     \\
\midrule
Euros  Consumption        & -0.409  & 0.468   & -0.147  & 0.369   & -0.672  & -0.065  \\
Total Energy Consumption      & -0.436  & -0.319  & -0.087  & 0.160   & 0.071   & 0.819   \\
Number of Active Males        & -0.373  & 0.651   & -0.185  & -0.231  & 0.591   & 0.030   \\
Electricity Consumption       & -0.426  & -0.344  & 0.041   & 0.562   & 0.368   & -0.498  \\
Coal and Derivatives Consumption & -0.397  & 0.004   & 0.849   & -0.326  & -0.115  & -0.046  \\
Gas Consumption               & -0.406  & -0.370  & -0.463  & -0.602  & -0.215  & -0.273  \\
\bottomrule
\end{tabular}
\end{table}

The Principal Component Analysis (PCA) results (see Table \ref{tab:PCA_results} reveal that the first three components explain approximately 97.8\% of the total variance in the selected covariates, with PC1 accounting for the majority (80.4\%), PC2 contributing 11.7\%, and PC3 adding 5.7\%. The remaining components collectively explain less than 2.5\% of the variance and were therefore excluded from the analysis.

PC1 appears to capture the overall economic activity and energy consumption patterns, as evidenced by strong negative loadings for total energy consumption, electricity consumption, and gas consumption, as well as the number of active males. This component likely reflects general production and consumption trends closely linked to GDP.

PC2 shows notable positive loadings on the number of active males and electricity consumption, and a negative loading on total energy consumption. This suggests it may represent structural changes in labor market participation and energy usage efficiency that can affect regional GDP dynamics differently than overall consumption.

PC3 is characterized by a strong positive loading on coal and derivatives consumption and a moderate negative loading on gas consumption. This component could be related to the energy mix and its transition effects, potentially capturing regional heterogeneities in industrial activity and energy sources relevant to GDP fluctuations.

Including the first three principal components balances the need to capture the dominant variation in the data while avoiding overfitting noise. Despite the smaller explained variance of PC3 compared to PC1 and PC2, it contains meaningful information about energy source composition and labor structure that is economically relevant for modeling regional GDP. The last three components, with negligible explained variance (less than 2.5\% combined), likely represent noise or idiosyncratic fluctuations and were excluded to maintain model parsimony and interpretability.

Thus, the selection of these three components ensures that the disaggregation model incorporates the main systematic factors affecting GDP across regions and time while filtering out less informative variation, which could otherwise deteriorate model stability and predictive performance.

The disaggregation methodology was then implemented using the first three principal components as predictors. The estimates of the model parameters are shown in Table \ref{tab:estimadores}.

\begin{table}[H]
\centering
\caption{Estimated parameters from the disaggregation model}
\label{tab:estimadores}
\begin{tabular}{lr}
\hline
Parameter & Estimate \\
\hline
$\rho$ & $0.6141$ \\
PC1 coefficient & $3.6125$ \\
PC2 coefficient & $3.9910$ \\
PC3 coefficient & $0.8781$ \\
$\sigma^2$ & $0.2707$ \\
$\phi_1$ & $0.8961$ \\
\hline
\end{tabular}
\end{table}

To assess model performance, we compared the observed and disaggregated GDP series using two metrics: Mean Absolute Percentage Error (MAPE) and Relative Root Mean Squared Error (RRMSE) relative to the mean GDP. These results are reported in Table \ref{tab:errors}. In addition to the Gower-based dissimilarity matrix, we also considered a conventional spatial contiguity matrix based on shared borders between regions. However, this alternative yielded consistently worse results in terms of both error metrics, reinforcing the advantage of the Gower approach for capturing interregional similarities in this application.

\begin{table}[H]
\centering
\caption{Disaggregation error metrics with and without anchoring}
\label{tab:errors}
\begin{tabular}{llcc}
\hline
$W$&Scenario & MAPE & RRMSE \\
\hline
Spatial neighboord&Without anchoring &  $0.8426$ & $0.6513$ \\
Spatial neighboord &With anchoring in 2002 &$0.2301$ & $0.2930$ \\
Gower &Without anchoring  & $0.1548$ & $0.1824$\\
Gower &With anchoring in 2002 & $0.1068$ & $0.1459$ \\
\hline
\end{tabular}
\end{table}

In the second scenario, anchoring was performed using the 17 observed GDP values for the year 2002. This information was used to improve identification and evaluate the effects of anchoring on the disaggregation estimates. The results show a clear improvement in both error metrics, reflecting the benefit of incorporating partial observed data at specific time points.

\begin{figure}[ht]
	\centering
	\includegraphics[width=14cm]{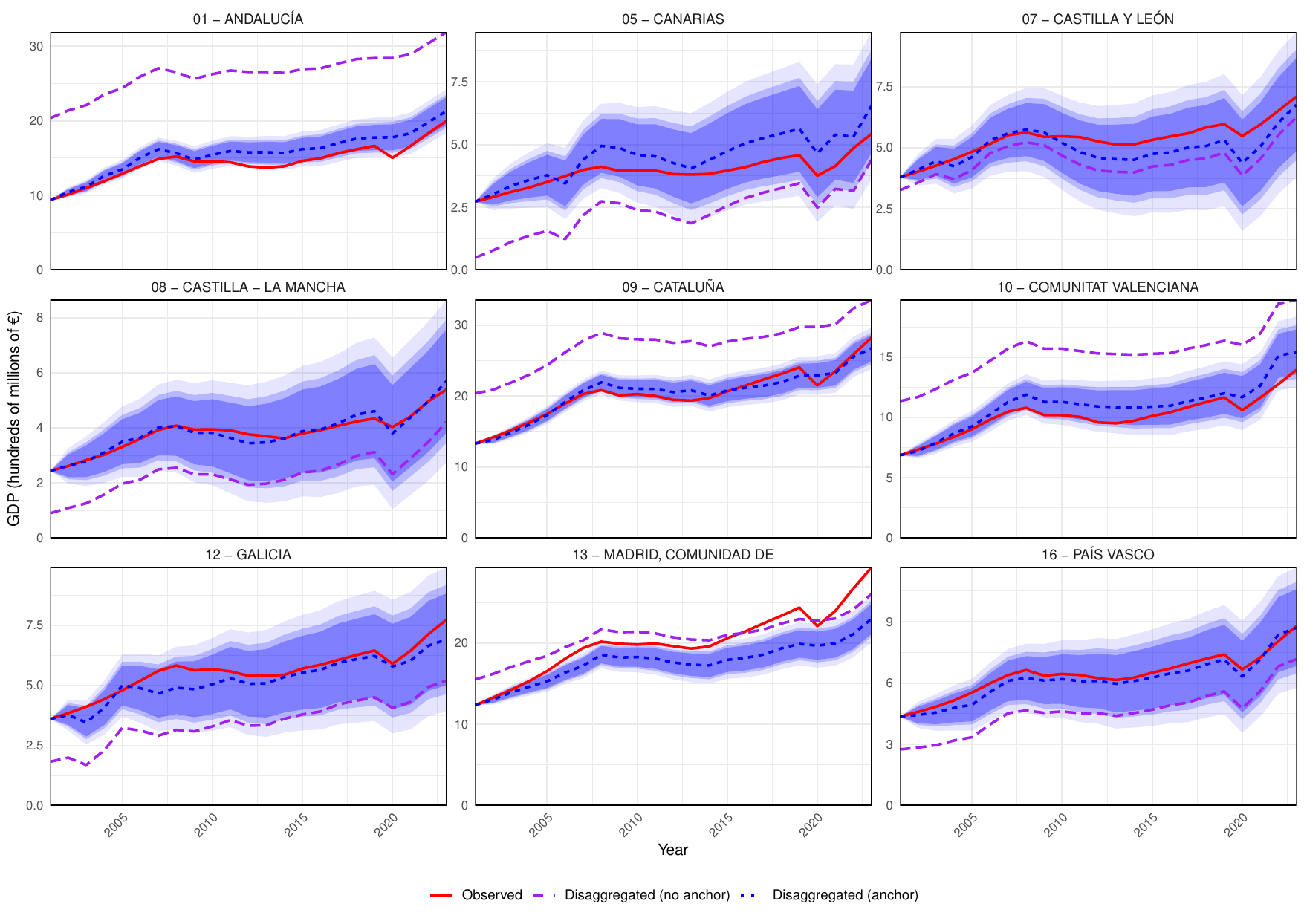}
	\caption{Evolution of GDP (in hundreds of millions of euros) from 2002 to 2023 for the nine Spanish autonomous communities with the highest average GDP, estimated using the proposed disaggregation methodology. The shaded bands represent pointwise confidence intervals at the 99\% (lightest), 95\%, and 90,\% (darkest) levels.}
	\label{fig:top9}
\end{figure}

\begin{figure}[ht]
	\centering
	\includegraphics[width=14cm]{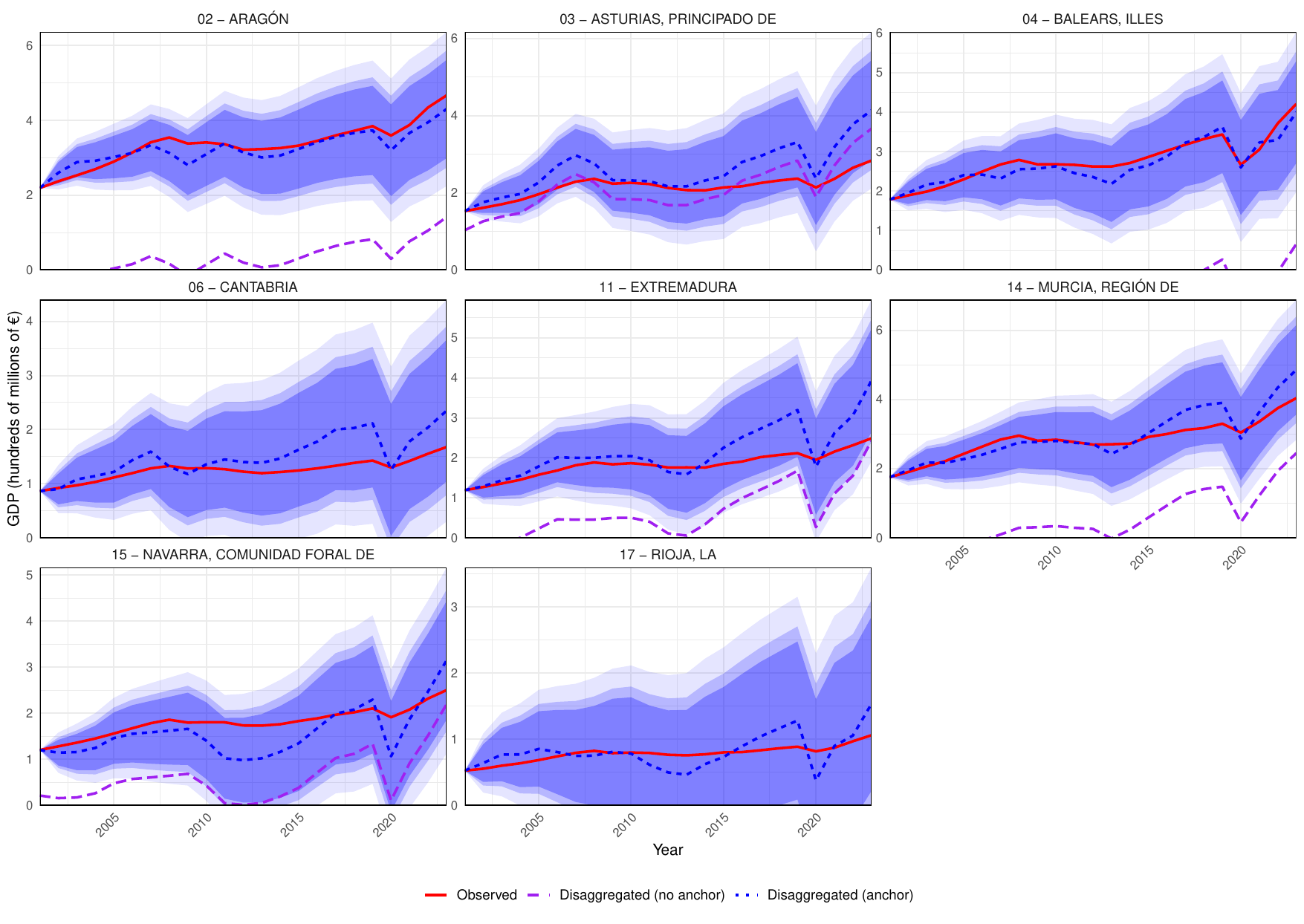}
	\caption{Evolution of GDP (in hundreds of millions of euros) from 2002 to 2023 for the nine Spanish autonomous communities with the highest average GDP, estimated using the proposed disaggregation methodology. The shaded bands represent pointwise confidence intervals at the 99\% (lightest), 95\%, and 90,\% (darkest) levels.}
	\label{fig:top8}
\end{figure}

\begin{table}[ht]
\centering
\begin{tabular}{ccccccc}
\hline
\textbf{CCAA} & \textbf{MAPE (\%)} & \textbf{RRMSE (\%)} & \textbf{R$^2$ (\%)} & \textbf{GDP} &\textbf{N-MAPE(\%)} \\
\hline
09 & 3.26  & 4.09  & 94.34 & 202.48 & 68.98 \\
13 & 11.03 & 14.55 & 96.81 & 200.97 & 68.56\\
01 & 7.90  & 9.36  & 95.59 & 144.68 & 56.64\\
10 & 7.54  & 9.59  & 95.61 & 101.33 & 38.16\\
16 & 4.15  & 4.52  & 97.68 & 63.85  & 1.83\\
12 & 6.65  & 8.37  & 90.09 & 55.69  & 12.80\\
07 & 7.99  & 9.84  & 73.86 & 53.40  & 17.02\\
05 & 15.02 & 18.04 & 89.58 & 39.27  & 59.43\\
08 & 3.06  & 4.24  & 95.17 & 38.18  & 64.67\\
02 & 5.87  & 7.51  & 85.92 & 33.52  & 87.26\\
14 & 7.89  & 12.14 & 89.15 & 28.28  & 122.82\\
04 & 6.08  & 7.43  & 89.09 & 27.60  & 128.94\\
03 & 18.66 & 26.40 & 84.22 & 21.57  & 189.90\\
11 & 18.06 & 28.50 & 81.85 & 18.22  & 244.69\\
15 & 17.63 & 22.99 & 61.91 & 18.04  & 247.72\\
06 & 20.52 & 28.45 & 77.24 & 12.52  & 399.75\\
17 & 20.23 & 27.06 & 40.35 & 7.82   & 700.91\\
\hline
\end{tabular}
\caption{Accuracy metrics of the disaggregated GDP estimates (with anchor) for each Spanish autonomous community, identified by CCAA code. MAPE = Mean Absolute Percentage Error; RRMSE = Relative Root Mean Squared Error; $R^2$ = Coefficient of Determination. GDP mean is expressed in millions of euros, N-MAPE = Mean Absolute Percentage Error of the Naive Estimator}
\label{tab:accuracy_metrics_ccaa}
\end{table}

The results of the disaggregation procedure applied to Spanish regional GDP reveal several important insights. First, Table~\ref{tab:errors} clearly shows that incorporating anchoring—by including the observed GDP values for the year 2002—significantly improves the accuracy of the estimates. Both MAPE and RRMSE are considerably reduced, demonstrating the stabilizing effect of anchoring in the reconstruction of regional series.

Figures~\ref{fig:top9} and~\ref{fig:top8} illustrate the evolution of the disaggregated GDP for the 17 autonomous communities, grouped by their average GDP level. It is apparent that regions with larger economies, such as \textit{Madrid}, \textit{Cataluña}, and \textit{Andalucía}, exhibit smoother and more stable trajectories. In contrast, regions with lower GDP, such as \textit{La Rioja}, \textit{Cantabria}, and \textit{Extremadura}, show greater variability and occasional deviations, even when anchoring is used. This pattern is further supported by Table~\ref{tab:accuracy_metrics_ccaa}, where we observe that the regions with higher average GDP tend to have lower MAPE and RRMSE, and higher $R^2$ values, indicating better predictive performance.

In addition, we computed the mean absolute percentage error (N-MAPE) of the naive estimator proposed in Equation~\eqref{naive}. 
The regional MAPEs reveal a poor adjustment to the observed data, with values ranging from $1.83\%$ to $700.91\%$, and a median above $60\%$. 
In particular, several regions display extremely large discrepancies, indicating that the block-averaging approach fails to reproduce the true disaggregated dynamics. 
This poor performance is consistent with the theoretical results derived in Theorem~\ref{the2}: when the conditions that make the naive and the proposed estimators equivalent are not simultaneously satisfied—specifically, when $\rho$ departs from $0$ and the regressors $\mathbf{Z}$ exhibit marked spatial heterogeneity—the proposed estimator achieves a substantially lower mean squared error than naive estimator.

The numerical results from the applied example strongly support the superiority of the proposed disaggregation estimator over the naive block-average predictor. In particular, the decomposition of
$\Delta$ in Theorem \ref{the2}
yields the following contributions (computed at the estimated parameter values): $
\pmb{\mu}^\top(\mathbb{I}_{nT}-\mathbf{P})\pmb{\mu} \approx 2.58\times 10^{4},\quad
2k_{\rho}\mathbf{m}^\top\mathbf{m} \approx 5.00,\quad
2\pmb{\mu}^\top(\mathbb{I}_{nT}-\mathbf{P})(\mathbb{I}_T\otimes\mathbf{D}_\rho)(\mathbf{Y}_a-\pmb{\mu}_a)\approx 42.8,$
while the parameter-uncertainty contribution is
$\operatorname{tr}(\mathbf{M}\operatorname{Var}(\hat\beta)\mathbf{M}^\top)\approx 273.4$.
The first term, which measures the spatial heterogeneity of the fitted systematic component $\pmb{\mu}=\mathbf{A}^{-1}\mathbf{Z}\pmb{\beta}$ not captured by block averaging, overwhelmingly dominates the decomposition. Consequently, $\Delta$ is large and positive, implying a substantial reduction in MSE when using the spatially informed predictor.

Note that, in the example, only one of the five conditions, high temporal persistence, $|\phi_1|\approx 1$) holds (approximately); the others do not: $\pmb{\beta}^\top\pmb{\beta}$ is not small, $\mathbf{Z}$ displays nontrivial within-period spatial variation ($\| \mathbf{Z}-\frac{1}{n}\mathbf{C}^\top\mathbf{CZ}\|\approx 1967$), and $\rho$ is moderate ($\rho\approx 0.614$), so the large gain is driven primarily by spatial heterogeneity in $\mathbf{A}^{-1}\mathbf{Z}\pmb{\beta}$. 

A particularly noteworthy finding is that the disaggregation approach based on only three principal components extracted from auxiliary variables was sufficient to explain the majority of the temporal and spatial variability in the regional GDP series. This dimensionality reduction preserved the key economic signals while discarding noise, making the model both efficient and interpretable. The effectiveness of this low-dimensional representation, combined with the spatial structure and the anchoring mechanism, is a central contribution of this work.

Overall, the proposed methodology offers a robust and flexible framework for temporal disaggregation in settings where only partial information is available. It succeeds in capturing both long-term economic trends and regional differences, while leveraging auxiliary information and limited direct observations to improve accuracy and interoperability.
\subsection*{Limitations}

Despite the promising results, the proposed disaggregation framework has some limitations. First, while anchoring improves estimation accuracy, its performance is highly dependent on the quality and representativeness of the observed reference year(s). If the anchored year corresponds to an atypical economic condition (e.g., recession or shock), this may introduce bias in the recovered trajectories.

Second, the current methodology assumes that the spatial weight matrix $W$ is fixed and known. Although using geographic contiguity or Gower distances provides reasonable structure, more sophisticated approaches could involve estimating or dynamically adjusting $W$ based on economic similarity or temporal changes, which was not explored in this work.

Furthermore, although the principal components capture most of the variation in auxiliary covariates, there is an implicit assumption that the relation between these covariates and the GDP is linear and stable over time. This may not hold in periods of structural economic change. Despite this limitation, Figures~\ref{fig:top9} and~\ref{fig:top8} show how the model correctly identifies the trends caused by the 2008 financial crisis and the 2020 COVID pandemic. Lastly, while the method offers strong empirical performance, further research could explore formal model selection strategies for determining the optimal number of components or penalization strength.

\section{Conclusions}
Beyond its empirical efficacy, the proposed disaggregation methodology is based on a rigorous theoretical framework. The estimation procedure is based on a linear model structure with a SAR model and an AR model in the error. This hybrid formulation allows for a fundamental balance between flexibility and identifiability.

Importantly, under standard regularity conditions and with an increasing time dimension $T$, the estimators of the model parameters are consistent and asymptotically normal, even in the absence of Gaussianity in the disturbances. This robustness to distributional assumptions constitutes a key theoretical advantage, especially relevant in economic applications where outliers or biased innovations are common.
The anchoring mechanism, modelled as a constrained estimation problem, guarantees identifiability and reduces variance while preserving consistency.
Together, these features make the method not only practically useful but also statistically reliable, combining interpretability, computational feasibility, and theoretical guarantees.

To empirically validate our proposal, a practical evaluation of the model has been performed. With the usage of synthetic data, results have shown how a controlled level of spatial correlation ($\rho$) permits efficient series data disaggregation. 

Furthermore, the problem of disaggregating the Spanish GDP by Autonomous Communities has been considered further to evaluate the validity of our model with real data.  In this scenario, the usage of PCA and anchoring shows how the usage of our model, in combination with common techniques in econometrics, permits a good disaggregation. 

As future work, we are interested in how to strengthen the model for values of $\rho$ closer to 1 and to further refine the capacity of the proposal under dynamic scenarios, that is, where the weight matrix $W$ is not fixed and when the relationship between covariates and the estimated value fluctuates over time.

\section*{Software implementation}
\section*{Supplementary files}
\begin{enumerate}
    \item Supplementary file 1\label{sf1}: Deduction of all conditions for naive estimator, 
    
    \item Supplementary file 2\label{sf2}: R files with application in GDP, 
\end{enumerate}

\appendix
\section{Matrices and Preliminary Results}\label{apenA0}

This appendix summarizes the main matrices, notation, and algebraic results used throughout the proofs. All matrices are assumed to have conformable dimensions.

\begin{align*}
\mathbf{1}_n &= (1,\ldots,1)^{\top}, &
\mathbf{J}_{n} &= \mathbf{1}_n\mathbf{1}_n^\top,\\
\mathbf{P} &= \mathbb{I}_{T} \otimes \frac{1}{n}\mathbf{J}_{n}, &
\mathbf{P}_{1} &= \mathbf{B}\mathbf{C}^\top \pmb{\Sigma}_{\mathbf{Y}_a}^{-1}\mathbf{C}, \\
\mathbf{F}& = \mathbb{I}_n - \rho \mathbf{W}, &
\mathbf{F}^{-1}&=\sum_{j=0}^\infty \rho^j \mathbf{W}^j,\\
 m_\rho &=\mathbf{1}_{n}^\top \mathbf{F}^{-1}(\mathbf{F}^{-1})^\top \mathbf{1}_{n}, &
\mathbf{D}_\rho &= \mathbf{F}^{-1}(\mathbf{F}^{-1})^\top m_\rho^{-1}\mathbf{1}_{n},\\
\mathbf{P}_0 &= \mathbb{I}_T\otimes\left(\mathbb{I}_n - m_\rho^{-1}\mathbf{F}^{-1}(\mathbf{F}^{-1})^\top\mathbf{J}_{n}\right),  &
\mathbf{P}_2 &= \pmb{\Sigma}_U^{-1}\otimes \frac{1}{n}\mathbf{J}_n,\\
\tilde{\mathbf{Z}}&=\mathbf{A}^{-1}\mathbf{Z}.
\end{align*}
The next properties will be used:
\begin{itemize}
    \item  If $\mathbf{H}$ is symmetric and idempotent matrix of $n$ columns, i.e, $\mathbf{H}\mathbf{H}=\mathbf{H}$, then $\mathbb{I}_n-\mathbf{H}$ and $\mathbb{I}_T\otimes \mathbf{H}$ are both symmetric and idempotent (Lemma 10.1.2 of \citep[pgs 134]{harville1998matrix}). $\mbox{rank}(\mathbb{I}_n-\mathbf{H})=n-\mbox{rank}(\mathbf{H})$ (Corollary 10.2.3 of \citep[pgs 135]{harville1998matrix})
    \item $\mbox{rank}(\mathbf{G}\otimes \mathbf{H})=\mbox{rank}(\mathbf{G})\mbox{rank}(\mathbf{H})$. (\citep[pgs 340-344]{harville1998matrix}).
    \item $$\mathbf{1}_n^\top \mathbf{H}\mathbf{1}_n=\mathbf{1}_n^\top \mathbf{H}^\top\mathbf{1}_n=\sum_{ij}\mathbf{H}_{ij}$$
\end{itemize}
By properties of the Kronecker product, the next results are true:
\begin{align}
\mathbf{B}&= \pmb{\Sigma}_U\otimes\mathbf{F}^{-1}(\mathbf{F}^{-1})^\top\\
\pmb{\Sigma}_{\mathbf{Y}_a} &=(\mathbb{I}_T\otimes \mathbf{1}_n^\top)\label{Kron1} (\pmb{\Sigma}_U\otimes\mathbf{F}^{-1}(\mathbf{F}^{-1})^\top)(\mathbb{I}_T\otimes \mathbf{1}_n)\\
&=\pmb{\Sigma}_U\otimes m_\rho =m_\rho\pmb{\Sigma}_U\label{ecsigmaAgre}\\
\pmb{\Sigma}_{\mathbf{Y}_a}^{-1} &=\frac{1}{m_\rho} \pmb{\Sigma}_U^{-1}\\
    \mathbf{P}_{1}&=\mathbf{BC}^\top \pmb{\Sigma}_{\mathbf{Y}_a}^{-1}\mathbf{C}\nonumber\\
    & =(\pmb{\Sigma}_U\otimes\mathbf{F}^{-1}(\mathbf{F}^{-1})^\top)(\mathbb{I}_T\otimes \mathbf{1}_n)( \pmb{\Sigma}_U^{-1}\otimes m_\rho^{-1} )(\mathbb{I}_T\otimes \mathbf{1}_n^\top)\nonumber\\
    &=\mathbb{I}_T\otimes \left(m_\rho^{-1}\mathbf{F}^{-1}(\mathbf{F}^{-1})^\top \mathbf{J}_{n}\right)\nonumber\\
    \mathbf{P}_{1}^\top &= \mathbb{I}_T\otimes\left( m_\rho^{-1}\mathbf{F}^{-1}(\mathbf{F}^{-1})^\top \mathbf{J}_{n}\right)^\top\nonumber\\
    &=\mathbb{I}_T\otimes \left( m_\rho^{-1}\mathbf{J}_{n}\mathbf{F}^{-1}(\mathbf{F}^{-1})^\top\right)\nonumber\\
    \mathbf{P}_{1}\mathbf{P}_{1}&=\mathbf{BC}^\top \pmb{\Sigma}_{\mathbf{Y}_a}^{-1}\mathbf{C}\mathbf{BC}^\top \pmb{\Sigma}_{\mathbf{Y}_a}^{-1}\mathbf{C}=\mathbf{BC}^\top \pmb{\Sigma}_{\mathbf{Y}_a}^{-1}\mathbf{C}=\mathbf{P}_{1}\\
    \mathbf{BC}^\top \pmb{\Sigma}_{\mathbf{Y}_a}^{-1} &= \mathbb{I}_T\otimes \mathbf{F}^{-1}(\mathbf{F}^{-1})^\top m_\rho^{-1}\mathbf{1}_{n} =  \mathbb{I}_T\otimes\mathbf{D}_\rho\label{Kronn}
\end{align}
Here, $\mathbf{P}$ and $\mathbf{P}_1$ are projection matrices, and therefore satisfy $\mathbf{P}^2=\mathbf{P}$, and $\mathbf{P}_1^2=\mathbf{P}_1$.

Since $\mathbf{W}$ is a stochastic matrix, let $c_w=\|\mathbf{W}\|_1$ denote its largest column sum.

The scalar quantity
\[
k_\rho = \mathbf{D}_\rho^\top\left(\mathbb{I}_n - \frac{1}{n}\mathbf{J}_n\right)\mathbf{D}_\rho \geq 0
\]
follows directly from the idempotence of $\mathbb{I}_n - \frac{1}{n}\mathbf{J}_n$.

Let $\lambda_l(\mathbf{G})$ denote the $l$-th eigenvalue of the matrix $\mathbf{G}$.

From von Neumann’s trace inequality \citep{mirsky1975trace} and the results of \citet[pp.~340–341]{marshall2011inequalities}, we obtain
\begin{align*}
\sum_{l=1}^k \lambda_l\!\left(\tilde{\mathbf{Z}}^\top\mathbf{P}_0^\top\mathbf{P}_0 
\tilde{\mathbf{Z}}\left[\tilde{\mathbf{Z}}^{\top}\mathbf{P}_2\tilde{\mathbf{Z}}\right]^{-1}\right)
&\leq 
\sum_{l=1}^k \lambda_l\!\left(\tilde{\mathbf{Z}}^\top\mathbf{P}_0^\top\mathbf{P}_0 \tilde{\mathbf{Z}}\right)
\lambda_l\!\left(\left[\tilde{\mathbf{Z}}^{\top}\mathbf{P}_2\tilde{\mathbf{Z}}\right]^{-1}\right).
\end{align*}

By the Rayleigh quotient theorem \citep[Theorem 4.2.2, p.~234]{horn2012matrix}, 
if $\mathbf{P}$ is a projection matrix, then for any vector $\mathbf{y}$,
\[
\|\mathbf{P}\mathbf{y}\|^2 \leq \|\mathbf{y}\|^2.
\]

Since $(\mathbb{I}_{nT}-\mathbf{P})$ is also a projection matrix, the standard properties hold:
\[
(\mathbb{I}_{nT}-\mathbf{P})^2 = \mathbb{I}_{nT}-\mathbf{P}, \quad
(\mathbb{I}_{nT}-\mathbf{P})^\top = \mathbb{I}_{nT}-\mathbf{P}.
\]
Hence, $\mathbb{I}_{nT}-\mathbf{P}$ projects onto the orthogonal complement of the column space of $\mathbf{P}$.
\begin{lemma}\label{lemamrho}
Let 
\[
m_\rho = \mathbf{1}_n^\top \mathbf{F}^{-1}(\mathbf{F}^{-1})^\top\mathbf{1}_n
= \big\|\mathbf{F}^{-T}\mathbf{1}_n\big\|^2.
\]
Then:
\begin{enumerate}
    \item $m_{\rho}$ is strictly increasing for all $\rho \in (0,1)$, and
    \item $m_{\rho}< m_{\vert \rho\vert }$ for $\rho<0$.
\end{enumerate}
\end{lemma}
\begin{proof}
By properties of the derivative of a matrix, we obtain that:
\begin{align*}
 m'_{\rho} =  \frac{\partial m_{\rho}}{\partial \rho} &= \frac{\partial}{\partial \rho}  \mathbf{1}_n^\top \mathbf{F}^{-1}(\mathbf{F}^{-1})^\top\mathbf{1}_n\\
    &= \mathbf{1}_n^\top \mathbf{F}^{-1}\mathbf{W}\mathbf{F}^{-1}(\mathbf{F}^{-1})^\top\mathbf{1}_n+\mathbf{1}_n^\top \mathbf{F}^{-1}(\mathbf{F}^{-1})^\top\mathbf{W}^\top(\mathbf{F}^{-1})^\top\mathbf{1}_n\\
    &=2\mathbf{1}_n^\top \mathbf{F}^{-1}\mathbf{W}\mathbf{F}^{-1}(\mathbf{F}^{-1})^\top\mathbf{1}_n
\end{align*}
and we obtain that $ m'_{0}= 2\mathbf{1}_n^\top \mathbb{I}_n\mathbf{W}\mathbb{I}_n(\mathbb{I}_n)^\top\mathbf{1}_n=2\mathbf{1}_n^\top \mathbf{W}\mathbf{1}_n=2\mathbf{1}_n^\top \mathbf{1}_n=2n$
Now, using the Neumann series expansion valid for $|\rho|<1$,
\begin{align*}
    \mathbf{F}^{-1}&= \sum_{j=0}^\infty \rho^j \mathbf{W}^j\\
       \mathbf{F}^{-1}\mathbf{F}^{-T} &= \left(\sum_{j=0}^\infty \rho^j \mathbf{W}^j \right)\left(\sum_{j=0}^\infty \rho^j (\mathbf{W}^j)^\top \right)\\+
       &=\sum_{j=0}^\infty\sum_{j'=0}^j\rho^{j'} \rho^{j-j'} \mathbf{W}^{j'} (\mathbf{W}^{j-j'})^\top \\
       &=\sum_{j=0}^\infty\rho^j\sum_{j'=0}^j\mathbf{W}^{j'} (\mathbf{W}^{j-j'})^\top \\
      \frac{\partial \mathbf{F}^{-1}\mathbf{F}^{-T}}{\partial \rho} &=  \sum_{j=1}^\infty j\rho^{j-1}\sum_{j'=0}^j\mathbf{W}^{j'} (\mathbf{W}^{j-j'})^\top \\
       \frac{\partial m_{\rho}}{\partial \rho} &=  \sum_{j=1}^\infty j\rho^{j-1}\sum_{j'=0}^j\mathbf{1}_n^\top\mathbf{W}^{j'} (\mathbf{W}^{j-j'})^\top \mathbf{1}_n\\
       &=\sum_{j=1}^\infty j\rho^{j-1}s_j
\end{align*}
with $s_j=\sum_{j'=0}^j\mathbf{1}_n^\top\mathbf{W}^{j'} (\mathbf{W}^{j-j'})^\top \mathbf{1}_n$. It is clear that $s_j>0$ for all $j$, hence, if $\rho>0$, the value $j\rho^{j-1}s_j>0$, and for $\rho'>\rho$ and $j>0$, $j(\rho')^j>j\rho^j>0$, and  it is clear that,
$$\sum_{j=0}^\infty j(\rho')^{j-1}s_j>\sum_{j=0}^\infty j\rho^{j-1}s_j>0.$$
Then we obtain that $0<2n<\frac{\partial m_{\rho}}{\partial \rho}$ for $\rho>0$. And for $\rho<0$
\begin{align*}
    m_{\rho} & = \sum_{j=0}^\infty \rho^{j}s_j <  \sum_{j=0}^\infty \vert \rho\vert ^{j}s_j = m_{\vert \rho \vert}
\end{align*}
\end{proof}

\begin{lemma}\label{Slustky}
    Let $\mathbf{X}_T$ an random vector of $(\mathbb{R}^k, \mathcal{B}(\mathbb{R}^k))$ and $\mathbf{F}_T$ a random non-singular matrix of $(\mathbb{R}^{k\times k}, \mathcal{B}(\mathbb{R}^{k\times k}))$. If $\mathbf{X}_T\xrightarrow[T\to\infty]{p}\mathbf{X}$  and $\mathbf{F}_T\xrightarrow[T\to\infty]{p}\mathbf{F}$ with $\mathbf{F}$ non-singular and constant matrix, then
    \begin{align*}
        \mathbf{F}_T\mathbf{X}_T\xrightarrow[T\to\infty]{d}\mathbf{F}\mathbf{X}\\
        \mathbf{F}^{-1}_T\mathbf{X}_T\xrightarrow[T\to\infty]{d}\mathbf{F}^{-1}\mathbf{X}
    \end{align*}
\end{lemma}
\begin{proof}
\begin{enumerate}
    \item Since $\mathbf{X}_T \xrightarrow[T\to\infty]{p} \mathbf{X}$ and $\mathbf{F}_T \xrightarrow[T\to\infty]{p} \mathbf{F}$, 
    we have 
    $(\mathbf{X}_T,\mathbf{F}_T)\xrightarrow[T\to\infty]{p}(\mathbf{X},\mathbf{F})$.    
   \item Consider any subsequence $\{n_t\}\subset\mathbb{N}$. Because $\mathbf{F}_T\xrightarrow[T\to\infty]{p}\mathbf{F}$ and 
    $\mathbf{X}_T\xrightarrow[T\to\infty]{p}\mathbf{X}$, by Theorem 6.3.1 b) of \citet[pg 172]{resnick2019probability}, exists a further subsequence $\{n_{t_\ell}\}\subset\{n_t\}$ 
    such that $\mathbf{F}_{n_{t_\ell}}\xrightarrow[T\to\infty]{a.s.}\mathbf{F}$ and $\mathbf{X}_{n_{t_\ell}}\xrightarrow[T\to\infty]{a.s.}\mathbf{X}$
    
    \item For each entry $i=1,\ldots,k$ of $\mathbf{F}_T\mathbf{X}_T$, we have
    \[
    (\mathbf{F}_T\mathbf{X}_T)_{i} = \sum_{r=1}^k (\mathbf{F}_{T})_{ir}\,(\mathbf{X}_{T})_{r}
    \]
    \item  By 2., for each fixed $r$, $(\mathbf{F}_{n_{t_\ell}})_{ir}(\mathbf{X}_{n_{t_\ell}})_r\xrightarrow[T\to\infty]{a.s.} (\mathbf{F})_{ir}(\mathbf{X})_{r}$.
    \item Hence, summing over $r=1,\ldots,k$, by Corollary 6.3.1 i) of \citet[pg 174]{resnick2019probability}, we have
    \[
    (\mathbf{F}_{n_{t_\ell}}\mathbf{X}_{n_{t_\ell}})_{i}\xrightarrow[T\to\infty]{a.s.}(\mathbf{F}\mathbf{X})_{i}.
    \]
    Since this holds for all $i$, we have
    \[
    \mathbf{F}_{n_{t_\ell}}\mathbf{X}_{n_{t_\ell}} \xrightarrow[T\to\infty]{a.s.} \mathbf{F}\mathbf{X}.
    \]
    \item And by Theorem 6.3.1 b) of \citet[pg 172]{resnick2019probability}, it is clear that:
      \[
    \mathbf{F}_{T}\mathbf{X}_{T} \xrightarrow[T\to\infty]{p} \mathbf{F}\mathbf{X}.
    \]
    \item Therefore, $\mathbf{F}_t\mathbf{X}_T\xrightarrow[T\to\infty]{p}\mathbf{F}\mathbf{X}$, then by Proposition 8.5.1 of \citet[pg 267]{resnick2019probability}, we have $\mathbf{F}_t\mathbf{X}_T\xrightarrow[T\to\infty]{d}\mathbf{F}\mathbf{X}$.
    \item How $\mathbf{F}_T\xrightarrow[T\to\infty]{p}\mathbf{F}$ and $\mathbf{F}^{-1}$ exists, 
    by Continous mapping Theorem (Theorem 2.3 of \citet[pg 7]{van2000asymptotic})$, \mathbf{F}^{-1}_T\xrightarrow[T\to\infty]{p}\mathbf{F}^{-1}$, and apllyng 1.-7. It is obtained that
    \[ \mathbf{F}^{-1}_T\mathbf{X}_T\xrightarrow[T\to\infty]{d}\mathbf{F}^{-1}\mathbf{X}\]
\end{enumerate}
\end{proof}

\section{Proof of Theorem \ref{the1}}\label{apenA}
\begin{proof}
\begin{enumerate}
\item Let two parameter values
\begin{align*}
\pmb{\theta}=(\pmb{\beta},\rho,\phi_1,\sigma^2),\qquad
\pmb{\theta}'=(\tilde{\pmb{\beta}},\tilde{\rho},\tilde{\phi}_1,\tilde{\sigma}^2)
\end{align*}
produce the same distribution for the observed aggregated vector $\mathbf{Y}_a$ (i.e. same Gaussian law). From Equation \eqref{ecsigmaAgre}, $\pmb{\Sigma}_Y=m_\rho\pmb{\Sigma}_U$. Let $\mathbf{R}(\phi_1)=\left(\frac{1-\phi_1^2}{\sigma^2}\right)\pmb{\Sigma}_U$, then it is clear that,
\begin{align*}
m_{\rho}\frac{\sigma^2}{1-\phi_1^2}\mathbf{R}(\phi_1)
=m_{\tilde{\rho}}\frac{\tilde{\sigma}^2}{1-\tilde{\phi}_1^2}\mathbf{R}(\tilde{\phi}_1).
\end{align*}
Denote the positive scalars
\begin{align*}
s=m_{\rho}\frac{\sigma^2}{1-\phi_1^2},\qquad
\tilde{s}=m_{\tilde{\rho}}\frac{\tilde{\sigma}^2}{1-\tilde{\phi}_1^2}.
\end{align*}
Then $s\mathbf{R}(\phi_1)=\tilde{s}\mathbf{R}(\tilde{\phi}_1)$.
Compare diagonal entries $(t,t)$ of $s\mathbf{R}(\phi_1)=\tilde{s} \mathbf{R}(\tilde{\phi}_1)$. Since $\mathbf{R}(\cdot)$ has ones on the diagonal,
\begin{align*}
s=\tilde{s}.
\end{align*}
Cancel the common scalar to obtain $\mathbf{R}(\phi_1)=\mathbf{R}(\tilde{\phi}_1)$. Inspect the first off-diagonal $(t,t+1)$: $[\mathbf{R}(\phi_1)]_{t,t+1}=\phi_1$ and $[\mathbf{R}(\tilde{\phi}_1)]_{t,t+1}=\tilde{\phi}_1$, so
\begin{align*}
\phi_1=\tilde{\phi}_1.
\end{align*}
Thus, the temporal AR(1) parameter is identified.

With $\phi_1=\tilde{\phi}_1$ and $s=\tilde{s}$ we have
\begin{align*}
m_{\rho}\frac{\sigma^2}{1-\phi_1^2}
=m_{\tilde{\rho}}\frac{\tilde{\sigma}^2}{1-\phi_1^2},
\end{align*}
hence
\begin{align*}
m_{\rho}\sigma^2=m_{\tilde{\rho}}\tilde{\sigma}^2.
\end{align*}
Therefore, the product $m_{\rho}\sigma^2$ is identified, but not yet the individual factors.
The mean is
\begin{align*}
\pmb{\mu}_a =\mathbf{C}\mathbf{A}^{-1}\mathbf{Z}\pmb{\beta}.
\end{align*}
Equality of distributions implies equality of means:
\begin{align*}
\mathbf{C}\mathbf{A}^{-1}\mathbf{Z}\pmb{\beta}=\mathbf{C}\tilde{\mathbf{A}}^{-1}\mathbf{Z}\tilde{\pmb{\beta}}.
\end{align*}
By defining $h(\rho, \pmb{\beta})=\mathbf{C}\mathbf{A}^{-1}\mathbf{Z}\pmb{\beta}$, the mapping $h:(-1,1)\times\mathbb{R}^k \to \mathbb{R}^T$ is continuously differentiable with Jacobian
\begin{align*}
    \mathcal{J}_{h}=\frac{\partial h(\rho, \pmb{\beta})}{\partial(\rho, \pmb{\beta})}&=\begin{pmatrix}
        \mathbf{C}\mathbf{A}^{-1}(\mathbb{I}_T\otimes\mathbf{W})\mathbf{A}^{-1}\mathbf{Z}\pmb{\beta}&
        \mathbf{C}\mathbf{A}^{-1}\mathbf{Z}
    \end{pmatrix}
\end{align*}
Under Assumption~\ref{as345}, $\mathrm{rank}(\mathcal{J}_{h})=k+1$, i.e., the Jacobian has full column rank.
Since the Jacobian $J_h(\rho,\pmb{\beta})$ has rank $k+1$, there exist $k+1$ components of $h$, denoted $h_{(1)},\ldots,h_{(k+1)}$, such that the submapping
\[
\tilde{h}(\rho,\pmb{\beta})
    = \bigl(h_{(1)}(\rho,\pmb{\beta}),\ldots,
       h_{(k+1)}(\rho,\pmb{\beta})\bigr)
\]
has full rank $k+1$. The mapping $\tilde{h}$ is one-to-one.
Therefore, the equation
\[
\tilde{h}(\rho,\pmb{\beta})
    = \tilde{h}(\tilde{\rho},\tilde{\pmb{\beta}}_0)
\]
admits a unique solution in $(-1,1)\times\mathbb{R}^k$, which establishes the identification of $(\rho,\pmb{\beta})$.

With $\rho=\tilde{\rho}$ known, $m_{\rho}$ is known. From $m_{\rho}\sigma^2=m_{\tilde{\rho}}\tilde{\sigma}^2$ we deduce $\sigma^2=\tilde{\sigma}^2$. 
All components of $\pmb{\theta}$ coincide with those of $\pmb{\theta}'$. Hence
\begin{align*}
(\pmb{\beta},\rho,\phi_1,\sigma^2)\longmapsto(\pmb{\mu}_a  ,\pmb{\Sigma}_Y)
\end{align*}
is injective, and the multivariate Gaussian likelihood of the aggregated data is identifiable.
\item 
Let $\ln L(\pmb{\theta})$ defined as:
\begin{equation}
     \ln L(\pmb{\theta})=-\frac{1}{2} \log |\pmb{\Sigma}_{\mathbf{Y}_a}\vert - \frac{1}{2} \left( \mathbf{Y}_a - \pmb{\mu}_a \right)^\top \pmb{\Sigma}_{\mathbf{Y}_a}^{-1} \left( \mathbf{Y}_a - \pmb{\mu}_a \right)\nonumber
\end{equation}
where
\begin{align*}
    \pmb{{\mu}}&=\mathbf{C} \mathbf{A}^{-1} \mathbf{Z} \pmb{\beta}\\
        \pmb{\Sigma}_{\mathbf{Y}_a} &= \mathbf{C} \mathbf{A}^{-1} (\pmb{\Sigma}_U \otimes \mathbb{I}_n) (\mathbf{A}^{-1})^{\top} \mathbf{C}^\top
    \end{align*}
Hence, the gradient of $\ln L(\pmb{\theta})$ with respect to $\pmb{\beta}$ is:
\begin{align}
   \frac{\partial \ln L(\pmb{\theta})}{\partial \pmb{\beta}} &=\left(\frac{\partial\pmb{\mu}_a}{\partial\pmb{\beta}} \right)^\top  \pmb{\Sigma}_{\mathbf{Y}_a}^{-1}\mathbb{E}=(\mathbf{Z}^\top(\mathbf{A}^{-1})^{\top}\mathbf{C}^\top)\pmb{\Sigma}_{\mathbf{Y}_a}^{-1}\mathbf{e}\label{deltabeta}\\
\mathbb{E}\left(\frac{\partial \ln L(\pmb{\theta})}{\partial \pmb{\beta}}\frac{\partial \ln L(\pmb{\theta})}{\partial \pmb{\beta}^\top}\right) &=(\mathbf{Z}^\top(\mathbf{A}^{-1})^{\top}\mathbf{C}^\top)\pmb{\Sigma}_{\mathbf{Y}_a}^{-1} \mathbf{C}\mathbf{A}^{-1}\mathbf{Z}\nonumber\\
&=-\mathbb{E}\left(\frac{\partial^2 \ln L(\pmb{\theta})}{\partial \pmb{\beta}\partial \pmb{\beta}^\top}\right)\label{fisherbeta}
\end{align}
Under the model in Equation \eqref{sart} and Assumptions \ref{as1}--\ref{as4}, write the stacked observation equation as
\[
\mathbf{Y}_a = \mathbf{C}\mathbf{A}^{-1}\mathbf{Z}\,\pmb{\beta} + \mathbf{e},
\]
with $\mathbb{E}(\mathbf{e})=\mathbf{0}$ and $\mathrm{Var}(\mathbf{e})=\pmb{\Sigma}_{\mathbf{Y}_a}$. The feasible GLS estimator for $\pmb{\beta}$ is \cite{bai2021feasible}:
\[
\hat{\pmb{\beta}} = \left[\mathbf{Z}^\top(\mathbf{A}^{-1})^\top\mathbf{C}^\top\widehat{\pmb{\Sigma}}_{\mathbf{Y}_a}^{-1}\mathbf{C}\mathbf{A}^{-1}\mathbf{Z}\right]^{-1}
\mathbf{Z}^\top(\mathbf{A}^{-1})^\top\mathbf{C}^\top\widehat{\pmb{\Sigma}}_{\mathbf{Y}_a}^{-1}\mathbf{Y}_a,
\]
where $\widehat{\pmb{\Sigma}}_Y$ is a  estimator of $\pmb{\Sigma}_{\mathbf{Y}_a}$ obtained from the estimated nuisance parameters.

When the true covariance $\pmb{\Sigma}_{\mathbf{Y}_a}$ is known, the estimator reduces to the infeasible GLS \cite{bai2021feasible}:
\begin{align*}
    \tilde{\pmb{\beta}}& = \left[\mathbf{Z}^\top(\mathbf{A}^{-1})^\top\mathbf{C}^\top\pmb{\Sigma}_{\mathbf{Y}_a}^{-1}\mathbf{C}\mathbf{A}^{-1}\mathbf{Z}\right]^{-1}
\mathbf{Z}^\top(\mathbf{A}^{-1})^\top\mathbf{C}^\top\pmb{\Sigma}_{\mathbf{Y}_a}^{-1}\mathbf{Y}_a\\
& =\left[\mathbf{Z}^\top(\mathbf{A}^{-1})^\top\mathbf{C}^\top\pmb{\Sigma}_{\mathbf{Y}_a}^{-1}\mathbf{C}\mathbf{A}^{-1}\mathbf{Z}\right]^{-1}
\mathbf{Z}^\top(\mathbf{A}^{-1})^\top\mathbf{C}^\top\pmb{\Sigma}_{\mathbf{Y}_a}^{-1}(\mathbf{C}\mathbf{A}^{-1}\mathbf{Z}\,\pmb{\beta} + \mathbf{e})\\
&=\left[\mathbf{Z}^\top(\mathbf{A}^{-1})^\top\mathbf{C}^\top\pmb{\Sigma}_{\mathbf{Y}_a}^{-1}\mathbf{C}\mathbf{A}^{-1}\mathbf{Z}\right]^{-1}
\mathbf{Z}^\top(\mathbf{A}^{-1})^\top\mathbf{C}^\top\pmb{\Sigma}_{\mathbf{Y}_a}^{-1}\mathbf{C}\mathbf{A}^{-1}\mathbf{Z}\,\pmb{\beta}+\\
&\left[\mathbf{Z}^\top(\mathbf{A}^{-1})^\top\mathbf{C}^\top\pmb{\Sigma}_{\mathbf{Y}_a}^{-1}\mathbf{C}\mathbf{A}^{-1}\mathbf{Z}\right]^{-1}
\mathbf{Z}^\top(\mathbf{A}^{-1})^\top\mathbf{C}^\top\pmb{\Sigma}_{\mathbf{Y}_a}^{-1}\mathbf{e}\\
&=\pmb{\beta}+\left[\mathbf{Z}^\top(\mathbf{A}^{-1})^\top\mathbf{C}^\top\pmb{\Sigma}_{\mathbf{Y}_a}^{-1}\mathbf{C}\mathbf{A}^{-1}\mathbf{Z}\right]^{-1}
\mathbf{Z}^\top(\mathbf{A}^{-1})^\top\mathbf{C}^\top\pmb{\Sigma}_{\mathbf{Y}_a}^{-1}\mathbf{e}
\end{align*}

Now decompose $\tilde{\pmb{\beta}}-\pmb{\beta}$:
\[
\tilde{\pmb{\beta}}-\pmb{\beta}
= \left[\mathbf{Z}^\top(\mathbf{A}^{-1})^\top\mathbf{C}^\top\pmb{\Sigma}_{\mathbf{Y}_a}^{-1}\mathbf{C}\mathbf{A}^{-1}\mathbf{Z}\right]^{-1}
\mathbf{Z}^\top(\mathbf{A}^{-1})^\top\mathbf{C}^\top\pmb{\Sigma}_{\mathbf{Y}_a}^{-1}\mathbf{e}.
\]

Under Assumption \ref{as1} and Assumptions \ref{as345}--\ref{as4}, $\pmb{\Sigma}_{\mathbf{Y}_a}^{-\frac{1}{2}}\mathbf{e} \sim \mathcal{N}_T(\mathbf{0}, \mathbb{I}_n)$, then
\[
\mathbf{Z}^\top(\mathbf{A}^{-1})^\top\mathbf{C}^\top\pmb{\Sigma}_{\mathbf{Y}_a}^{-1}\mathbf{e}
\sim \mathcal{N}\big(\mathbf{0},\; \mathbf{Z}^\top(\mathbf{A}^{-1})^\top\mathbf{C}^\top\pmb{\Sigma}_{\mathbf{Y}_a}^{-1}\mathbf{C}\mathbf{A}^{-1}\mathbf{Z}\big).
\]
Multiplying on the left by the inverse matrix yields
\[
\tilde{\pmb{\beta}} \sim\mathcal{N}_k\big(\pmb{\beta},\; \big[\mathbf{Z}^\top(\mathbf{A}^{-1})^\top\mathbf{C}^\top\pmb{\Sigma}_{\mathbf{Y}_a}^{-1}\mathbf{C}\mathbf{A}^{-1}\mathbf{Z}\big]^{-1}\big).
\]
Finally, to pass from the infeasible GLS $\tilde{\pmb{\beta}}$ to the feasible estimator $\hat{\pmb{\beta}}$ (FGLS), note  that under Assumptions \ref{as1}--\ref{as3} and by Corollary 3.8 of \citet[pg 448]{lehmann1998theory}, $\widehat{\pmb{\Sigma}}_Y$ is a consistent  estimator of $\pmb{\Sigma}_{\mathbf{Y}_a}$, and the matrix
\[
\mathbf{Z}^\top(\mathbf{A}^{-1})^\top\mathbf{C}^\top\widehat{\pmb{\Sigma}}_Y^{-1}\mathbf{C}\mathbf{A}^{-1}\mathbf{Z}
\]
converges in probability to its probability limit
\[
\mathbf{Z}^\top(\mathbf{A}^{-1})^\top\mathbf{C}^\top\pmb{\Sigma}_{\mathbf{Y}_a}^{-1}\mathbf{C}\mathbf{A}^{-1}\mathbf{Z},
\]
which is nonsingular by Assumption \ref{as345}.
Let
\[\mathbf{F}_T = \mathbf{Z}^\top(\mathbf{A}^{-1})^\top\mathbf{C}^\top\widehat{\pmb{\Sigma}}_Y^{-1}\mathbf{C}\mathbf{A}^{-1}\mathbf{Z}\]
\[\mathbf{X}_T =\mathbf{Z}^\top(\mathbf{A}^{-1})^\top\mathbf{C}^\top\pmb{\Sigma}_{\mathbf{Y}_a}^{-1}\mathbf{e},\]
therefore by Theorem \ref{Slustky}, 
\[
\hat{\pmb{\beta}} \xrightarrow[T\to\infty]{d} \mathcal{N}_k\big(\pmb{\beta},\; \big[\mathbf{Z}^\top(\mathbf{A}^{-1})^\top\mathbf{C}^\top\pmb{\Sigma}_{\mathbf{Y}_a}^{-1}\mathbf{C}\mathbf{A}^{-1}\mathbf{Z}\big]^{-1}\big).
\]
Multiplying out the inverse in the displayed variance gives the variance matrix in Equation \eqref{normalte1} of the theorem.

\item Using Theorem 8.22 of \citet[pg 62]{lehmann1998theory}, i) and ii) it is clear that iii) is satisfied.
 \end{enumerate}
\end{proof}
\section{Proof of Theorem \ref{the2}}\label{AppenB}
\begin{proof}
Let the naive predictor of $\mathbf{Y}$ be defined as
\[
\hat{\mathbf{Y}}^{1} = \frac{1}{n}\mathbf{C}^\top\mathbf{Y}_a=\mathbf{P}\mathbf{Y}, \qquad 
\mathbf{P} = \mathbb{I}_{T} \otimes \frac{1}{n}\mathbf{J}_{n},
\]
where $\mathbf{J}_{n}=\mathbf{1}_n\mathbf{1}_n^\top$ denotes the $n \times n$ matrix of ones and $\mathbb{I}_{T}$ the $T$-dimensional identity matrix. This estimator corresponds to the temporal vector of spatial means, obtained by averaging observations within each time period.

Since $\mathbf{Y} = \pmb{\mu} + \mathbf{e}$ with $ \mathbf{e}=\mathbf{A}^{-1}\mathbf{U}$,  $\mathbb{E}(\mathbf{e}) = \mathbf{0}$ and $\mathrm{Var}(\mathbf{e}) = \mathbf{B}$, the expectation of $\hat{\mathbf{Y}}^{1}\mid \mathbf{Y}_a$ is $$\mathbb{E}(\hat{\mathbf{Y}}^{1}\mid\mathbf{Y}_a) = \mathbf{P}(\pmb{\mu}+\mathbf{BC}^\top \pmb{\Sigma}_{\mathbf{Y}_a}^{-1}(\mathbf{Y}_a-\mathbf{C}\pmb{\mu})),$$ 
and using Equation \eqref{CondNormal}, the bias of the predictor is therefore
\begin{align}
  \mathrm{Bias}(\hat{\mathbf{Y}}^{1}\mid \mathbf{Y}_a) &= \mathbb{E}(\hat{\mathbf{Y}}^{1}\mid \mathbf{Y}_a) - \mathbb{E}(\mathbf{Y}\mid \mathbf{Y}_a)\nonumber\\
&= \mathbf{P}(\pmb{\mu}+\mathbf{BC}^\top \pmb{\Sigma}_{\mathbf{Y}_a}^{-1}(\mathbf{Y}_a-\mathbf{C}\pmb{\mu}))-\pmb{\mu}-\mathbf{BC}^\top\pmb{\Sigma}_{\mathbf{Y}_a}^{-1}(\mathbf{Y}_a-\pmb{\mu}_a)\nonumber \\
&=-(\mathbb{I}_{nT}-\mathbf{P})\left(\pmb{\mu}+\mathbf{BC}^\top \pmb{\Sigma}_{\mathbf{Y}_a}^{-1}(\mathbf{Y}_a-\mathbf{C}\pmb{\mu})\right)\nonumber\\
&=-(\mathbb{I}_{nT}-\mathbf{P})\pmb{\mu_{1a}}\label{muses1}
\end{align}
where $\pmb{\mu}_{1a}=\pmb{\mu}+\mathbf{BC}^\top \pmb{\Sigma}_{\mathbf{Y}_a}^{-1}(\mathbf{Y}_a-\mathbf{C}\pmb{\mu})$.
The corresponding covariance matrix of the prediction error is
\begin{align}
\mathrm{Var}(\mathbf{Y}-\hat{\mathbf{Y}}^{1}\mid\mathbf{Y}_a) 
&= (\mathbb{I}_{nT}-\mathbf{P})\,(\mathbf{B}-\mathbf{BC}^\top \pmb{\Sigma}_{\mathbf{Y}_a}^{-1}\mathbf{CB})\,(\mathbb{I}_{nT}-\mathbf{P})\nonumber\\
&=(\mathbb{I}_{nT}-\mathbf{P})(\mathbb{I}_{nT}-\mathbf{P}_{1})\mathbf{B}(\mathbb{I}_{nT}-\mathbf{P})\label{muses2}
\end{align}
where $\mathbf{P}_{1}=\mathbf{BC}^\top \pmb{\Sigma}_{\mathbf{Y}_a}^{-1}\mathbf{C}$.
Furthermore, the matrix $\mathbf{A}^{-1}$ can be expressed as:
\begin{align*}
    \mathbf{A}^{-1}&=(\mathbb{I}_T\otimes (\mathbb{I}_{n}-\rho  \mathbf{W}))^{-1}\\
    &=\mathbb{I}_T\otimes (\mathbb{I}_{n}-\rho  \mathbf{W})^{-1},
\end{align*}
then, let 
Hence combining Equations \eqref{muses1} and \eqref{muses2}, the mean squared error (MSE) of $\hat{\mathbf{Y}}^{1}$ with respect to $\mathbf{Y}$, and their expectations is given by:
\begin{align*}
\mathrm{MSE}&(\hat{\mathbf{Y}}^{1},\mathbf{Y})= \frac{1}{nT}\sum_{i=1}^{nT}  (\hat{Y}^{(1)}_{i}-Y_{i})^2\\
\mathbb{E}(\mathrm{MSE}&(\hat{\mathbf{Y}}^{1},\mathbf{Y})\mid \mathbf{Y}_a) = \frac{1}{nT}\sum_{i=1}^{nT} \mathbb{E}((\hat{Y}^{(1)}_{i}-Y_{i})^2\mid \mathbf{Y}_a)\\
 &=\frac{1}{nT}\sum_{i=1}^{nT} \mathbb{E}((\hat{Y}^{(1)}_{i}-Y_{i})^2\mid \mathbf{Y}_a)\\
 &=\frac{1}{nT}\sum_{i=1}^{nT} \left(\mbox{Var}((\hat{Y}^{(1)}_{i}-Y_{i})\mid \mathbf{Y}_a)+\mathbb{E}((\hat{Y}^{(1)}_{it}-Y_{it})\mid \mathbf{Y}_a)^2\right)\\
 &=\frac{1}{nT}\left(\mbox{tr}(\mbox{Var}(\mathbf{Y}-\hat{\mathbf{Y}}^{1}\mid\mathbf{Y}_a))+\mbox{Bias}(\hat{\mathbf{Y}}^{1}\mid \mathbf{Y}_a)\right)\\
&= \|(\mathbb{I}_{nT}-\mathbf{P})\pmb{\mu}_{1a}\|^{2}
+ \operatorname{tr}\big((\mathbb{I}_{nT}-\mathbf{P})(\mathbb{I}_{nT}-\mathbf{P}_{1})\mathbf{B}(\mathbb{I}_{nT}-\mathbf{P})\big)\\
&= \|(\mathbb{I}_{nT}-\mathbf{P})\pmb{\mu}_{1a}\|^{2}
+ \operatorname{tr}\big((\mathbb{I}_{nT}-\mathbf{P})(\mathbb{I}_{nT}-\mathbf{P}_{1})\mathbf{B}\big).
\end{align*}
The first term represents the squared bias due to ignoring the systematic component $\pmb{\mu}$, while the second term accounts for the variance of the residual component.

For comparison, for the estimator proposed in Equation \eqref{ecdesagrega} $\mathbf{Y}$ given $\mathbf{Y}_{a}$, using property in Equation \eqref{CondNormal} and Theorem \ref{the1}, it is clear that $\mathbb{E}(\hat{\mathbf{Y}}\mid \mathbf{Y}_a)=\mathbf{Y}$, and their mean squared error is
\begin{align*}
\mathbb{E}(\mathrm{MSE}(\hat{\mathbf{Y}},\mathbf{Y})\mid \mathbf{Y}_a) &= \mbox{tr}(\mbox{Var}(\hat{\mathbf{Y}}\mid \mathbf{Y}_a))\\
&= \operatorname{tr}\big(
\mathbf{B} - 
\mathbf{B}\mathbf{C}^{\top}(\mathbf{C}\mathbf{B}\mathbf{C}^{\top})^{-1}
\mathbf{C}\mathbf{B}
+\mathbf{M}\mbox{Var}(\hat{\pmb{\beta}})\mathbf{M}^\top\big)\\
&= \operatorname{tr}((\mathbb{I}_{nT}-\mathbf{P}_{1})\mathbf{B}) +\mbox{tr}(\mathbf{M}\mbox{Var}(\hat{\pmb{\beta}})\mathbf{M}^\top)
\end{align*}
where
\begin{align*}
    \mathbf{M}&=\mathbf{A}^{-1}\mathbf{Z}-\mathbf{B}\mathbf{C}^\top \pmb{\Sigma}_{\mathbf{Y}_a}^{-1}\mathbf{CA}^{-1}\mathbf{Z}=(\mathbb{I}_{nT}-\mathbf{P}_{1})\mathbf{A}^{-1}\mathbf{Z}\\
    &=\left(\mathbb{I}_T\otimes[\mathbb{I}_n-m_\rho^{-1}\mathbf{F}^{-1}(\mathbf{F}^{-1})^\top \mathbf{J}_{n}]\right)\mathbf{A}^{-1}\mathbf{Z}= \mathbf{P}_0\tilde{\mathbf{Z}}\\
    \mathbf{M}^\top \mathbf{M} &= \tilde{\mathbf{Z}}^\top\mathbf{P}_0^\top\mathbf{P}_0 \tilde{\mathbf{Z}}
\end{align*}
From Theorem \ref{the1} we obtain that
\begin{align*}
   \mbox{Var}(\hat{\pmb{\beta}})&= \left[\mathbf{Z}^\top(\mathbf{A}^{-1})^{\top}\mathbf{C}^\top\pmb{\Sigma}_{\mathbf{Y}_a}^{-1} \mathbf{C}\mathbf{A}^{-1}\mathbf{Z}\right]^{-1}\\
&=\left[\tilde{\mathbf{Z}}^{\top}\left(\pmb{\Sigma}_U^{-1}\otimes \frac{1}{m_\rho}\mathbf{J}_n\right)\tilde{\mathbf{Z}}\right]^{-1}\\
&=\left[\tilde{\mathbf{Z}}^{\top}\left(\frac{n}{m_\rho}\pmb{\Sigma}_U^{-1}\otimes \frac{1}{n}\mathbf{J}_n\right)\tilde{\mathbf{Z}}\right]^{-1}\\
&=\left[\tilde{\mathbf{Q}}^{\top}\left(\frac{n}{m_\rho}\pmb{\Sigma}_U^{-1}\otimes\mathbb{I}_n\right)\tilde{\mathbf{Q}}\right]^{-1}\\
&=\frac{m_\rho}{n}\left[\tilde{\mathbf{Q}}^{\top}(\pmb{\Sigma}_U^{-1}\otimes\mathbb{I}_n)\tilde{\mathbf{Q}}\right]^{-1}
=\frac{m_\rho}{n}\left[\tilde{\mathbf{Z}}^{\top}\mathbf{P}_2\tilde{\mathbf{Z}}\right]^{-1}\\
\end{align*}
where $\mathbf{P}_2=\pmb{\Sigma}_U^{-1}\otimes \frac{1}{n}\mathbf{J}_n=\pmb{\Sigma}_U^{-1}\otimes \frac{1}{\sqrt{n}}\mathbf{1}_n\frac{1}{\sqrt{n}}\mathbf{1}_n^\top=\mathbf{P}_2^\top$, and $\tilde{\mathbf{Q}}=(\mathbb{I}_T\otimes\frac{1}{\sqrt{n}}\mathbf{1}_n^\top)\tilde{\mathbf{Z}}$.

Then we obtain that $\mbox{rank}(\mathbf{P}_2)=T$, $\mbox{rank}(\mathbf{P}_0)=T(n-1)$, $\mbox{rank}(\tilde{\mathbf{Z}}^{\top})=\mbox{rank}(\tilde{\mathbf{Z}})=k$.

By Theorem 21.8.2 of \citet[pg 548]{harville1998matrix}, $\lambda_l(\mathbf{P}_0)=1$ if $1\leq l \leq T(n-1)$ and $\lambda_l(\mathbf{P}_0)=0$ if $l>T(n-1)$. 

Since $\mathbf{P}_0$ is idempotent, it is clear that $0\leq \lambda_l(\mathbf{P}_0^\top\mathbf{P}_0)\leq 1$, $1\leq l\leq nT$. $\lambda_1(\frac{1}{n}\mathbf{J}_n)=1$, and $\lambda_l(\frac{1}{n}\mathbf{J}_n)=0$, for $1<l\leq n$, and using Theorem 21.11.1 of \citep[pg 554]{harville1998matrix} we obtain that $\lambda_l(\mathbf{P}_2)=\lambda_l(\pmb{\Sigma}_U^{-1})\lambda_{l'}(\frac{1}{n}\mathbf{J}_n)$ for $1\leq l\leq T$, and $1\leq l'\leq n$.  

By Assumption \ref{as345}, $\lambda_l\left(\tilde{\mathbf{Z}}^\top\mathbf{P}_0^\top\mathbf{P}_0 \tilde{\mathbf{Z}}\right)\leq 0$, and $\lambda_l\left(\tilde{\mathbf{Z}}^\top\mathbf{P}_2 \tilde{\mathbf{Z}}\right)>0$ for $1\leq l\leq k$. 

Using \cite{10208005}, the eigenavalues of $\pmb{\Sigma}_U$ are given by
\begin{align*}
    \lambda_l(\pmb{\Sigma}_U) &= \frac{\sigma^2}{1-2\phi_1\cos{\left(\frac{l\pi}{T+1}\right)}+\phi_1^2}
\end{align*}
Hence, the largest eigenvalue is
\[
\lambda_{\max}(\pmb{\Sigma}_U)
= 
\frac{\sigma^2}{1 - 2|\phi_1|\cos\!\left(\frac{\pi}{T+1}\right) + \phi_1^2}
\geq 
\frac{\sigma^2}{(1 - |\phi_1|)^2},
\]
where the last approximation holds for large $T$, since 
$\cos\!\left(\frac{\pi}{T+1}\right) \approx 1$.
Hence, $\lambda_{\min}(\pmb{\Sigma}_U^{-1})\leq\frac{(1-\vert\phi_1 \vert)^2}{\sigma^2}=s_1$ then, using Corollary 4.3.15 of \citet[pg 242]{horn2012matrix} we obtain that 
\begin{align*}
    \lambda_l\left(\tilde{\mathbf{Z}}^\top \tilde{\mathbf{Z}}\right)&= \lambda_l\left(\tilde{\mathbf{Z}}^\top\mathbf{P}_0^\top\mathbf{P}_0 \tilde{\mathbf{Z}}+\tilde{\mathbf{Z}}^\top(\mathbb{I}_{nT}-\mathbf{P}_0^\top\mathbf{P}_0) \tilde{\mathbf{Z}}\right)\\
    &\geq \lambda_l\left(\tilde{\mathbf{Z}}^\top\mathbf{P}_0^\top\mathbf{P}_0 \tilde{\mathbf{Z}}\right)+ \lambda_{\min}\left(\tilde{\mathbf{Z}}^\top(\mathbb{I}_{nT}-\mathbf{P}_0^\top\mathbf{P}_0) \tilde{\mathbf{Z}}\right)\\
    & \geq  \lambda_l\left(\tilde{\mathbf{Z}}^\top\mathbf{P}_0^\top\mathbf{P}_0 \tilde{\mathbf{Z}}\right)
\end{align*}
\begin{align*}
\lambda_l\left(\left[\tilde{\mathbf{Z}}^{\top}\mathbf{P}_2\tilde{\mathbf{Z}}\right]^{-1}\right)&=\lambda_l\left(\left[\tilde{\mathbf{Q}}^{\top}\left(\pmb{\Sigma}_U^{-1}\otimes \mathbb{I}_n\right)\tilde{\mathbf{Q}}\right]^{-1}\right)\\
&=\frac{1}{s_1}\lambda_l\left(\left[\tilde{\mathbf{Q}}^{\top}\frac{1}{s_1}\left(\pmb{\Sigma}_U^{-1}\otimes \mathbb{I}_n\right)\tilde{\mathbf{Q}}\right]^{-1}\right)\\
&=\frac{1}{s_1\lambda_l\left(\tilde{\mathbf{Q}}^{\top}\frac{1}{s_1}\left(\pmb{\Sigma}_U^{-1}\otimes \mathbb{I}_n\right)\tilde{\mathbf{Q}}\right)}\\
&=\frac{1}{s_1\lambda_l\left(\tilde{\mathbf{Q}}^{\top}\tilde{\mathbf{Q}}+\tilde{\mathbf{Q}}^{\top}\left[\frac{1}{s_1}\left(\pmb{\Sigma}_U^{-1}\otimes \mathbb{I}_n\right)-\mathbb{I}_{nT}\right]\tilde{\mathbf{Q}}\right)}\\
&\leq \frac{1}{s_1\lambda_l\left(\tilde{\mathbf{Q}}^{\top}\tilde{\mathbf{Q}}\right)+s_1\lambda_{\min}\left(\tilde{\mathbf{Q}}^{\top}\left[\frac{1}{s_1}\left(\pmb{\Sigma}_U^{-1}\otimes \mathbb{I}_n\right)-\mathbb{I}_{nT}\right]\tilde{\mathbf{Q}}\right)}\\
&\leq \frac{1}{s_1\lambda_l\left(\tilde{\mathbf{Q}}^{\top}\tilde{\mathbf{Q}}\right)}=\frac{1}{s_1}\lambda_l\left(\left[\tilde{\mathbf{Q}}^{\top}\tilde{\mathbf{Q}}\right]^{-1}\right)
\end{align*}
And using the results above and von Neumann inequality \citep{mirsky1975trace}, we obtain that:
\begin{align}
    \mbox{tr}\left(\mathbf{M}\mbox{Var}(\hat{\pmb{\beta}})\mathbf{M}^\top\right)&=\mbox{tr}\left(\mbox{Var}(\hat{\pmb{\beta}})\mathbf{M}^\top\mathbf{M}\right)\nonumber\\
    &=\frac{m_\rho}{n}\mbox{tr}\left(\tilde{\mathbf{Z}}^\top\mathbf{P}_0^\top\mathbf{P}_0 \tilde{\mathbf{Z}}\left[\tilde{\mathbf{Z}}^{\top}\mathbf{P}_2\tilde{\mathbf{Z}}\right]^{-1}\right)\nonumber\\
&=\frac{m_\rho}{n}\sum_{l=1}^k\lambda_l\left(\tilde{\mathbf{Z}}^\top\mathbf{P}_0^\top\mathbf{P}_0 \tilde{\mathbf{Z}}\left[\tilde{\mathbf{Z}}^{\top}\mathbf{P}_2\tilde{\mathbf{Z}}\right]^{-1}\right)\nonumber\\
& \text{using \citet[pg 340-341]{marshall2011inequalities}}\nonumber\\
    &\leq \frac{m_\rho}{n} \sum_{l=1}^k\lambda_l\left(\tilde{\mathbf{Z}}^\top\mathbf{P}_0^\top\mathbf{P}_0 \tilde{\mathbf{Z}}\right)\lambda_l\left(\left[\tilde{\mathbf{Z}}^{\top}\mathbf{P}_2\tilde{\mathbf{Z}}\right]^{-1}\right)\nonumber\\
    &\leq \frac{m_\rho}{n} \sum_{l=1}^k\lambda_l\left(\tilde{\mathbf{Z}}^\top \tilde{\mathbf{Z}}\right)\frac{1}{s_1}\lambda_l\left(\left[\tilde{\mathbf{Q}}^{\top}\tilde{\mathbf{Q}}\right]^{-1}\right)\nonumber\\
    &=\frac{m_\rho}{ns_1}\sum_{l=1}^k\lambda_l\left(\tilde{\mathbf{Z}}^\top \tilde{\mathbf{Z}}\right)\lambda_l\left(\left[\tilde{\mathbf{Q}}^{\top}\tilde{\mathbf{Q}}\right]^{-1}\right)\label{trace1}
\end{align}
Let $\mathbf{P}=\mathbb{I}_T\otimes \frac{1}{n}\mathbf{J}_n$, we obtain that $\mathbf{P}=\mathbf{P}\mathbf{P}=\mathbf{P}^\top$ and $\mathbf{0}\neq \pmb{\nu} \in \mathbb{R}^k$, we obtain that
\begin{align}
    c_1&=\min_{\mathbf{0}\neq\mathbf{y}\in \langle col(\tilde{\mathbf{Z}})\rangle} \frac{\|\mathbf{P}\mathbf{y}\|^2}{\|\mathbf{y}\|^2}\nonumber\\
    \|\mathbf{P}\mathbf{y}\|^2&\geq c \|\mathbf{y}\|^2\; \text{ for }\mathbf{0}\neq\mathbf{y}\in \langle col(\tilde{\mathbf{Z}})\rangle\nonumber\\
    \pmb{\nu}^\top\tilde{\mathbf{Q}}^{\top}\tilde{\mathbf{Q}}\pmb{\nu}&=\|\mathbf{P}\tilde{\mathbf{Z}}\pmb{\nu}\|^2 \geq c_1 \|\tilde{\mathbf{Z}}\pmb{\nu}\|^2=c_1\pmb{\nu}^\top\tilde{\mathbf{Z}}^{\top}\tilde{\mathbf{Z}}\pmb{\nu}\label{ecraleuhg}
\end{align}
because $\tilde{\mathbf{Z}}\pmb{\nu}\in  col(\tilde{\mathbf{Z}})$. 
By Rayleigh Quotient as in Theorem 4.2.2 \citep[pg 234]{horn2012matrix}
\begin{align}
\frac{\pmb{\nu}^\top\tilde{\mathbf{Q}}^{\top}\tilde{\mathbf{Q}}\pmb{\nu}}{\pmb{\nu}^\top\pmb{\nu}} &\geq 
 \min_{1\leq l\leq  k}\lambda_l\left(\tilde{\mathbf{Q}}^{\top}\tilde{\mathbf{Q}}\right)\nonumber\\
\frac{\pmb{\nu}^\top\tilde{\mathbf{Z}}^{\top}\tilde{\mathbf{Z}}\pmb{\nu}}{\pmb{\nu}^\top\pmb{\nu}} &\geq 
 \min_{1\leq l\leq  k}\lambda_l\left(\tilde{\mathbf{Z}}^{\top}\tilde{\mathbf{Z}}\right)\geq c_1 \min_{1\leq l\leq  k}\lambda_l\left(\tilde{\mathbf{Z}}^{\top}\tilde{\mathbf{Z}}\right)\label{ecraleigh}
\end{align}
and replacing Equation \eqref{ecraleigh} in Equation \eqref{ecraleuhg} we obtain that 
\begin{align*}
\pmb{\nu}^\top\tilde{\mathbf{Q}}^{\top}\tilde{\mathbf{Q}}\pmb{\nu}&\geq c_1\pmb{\nu}^\top\tilde{\mathbf{Z}}^{\top}\tilde{\mathbf{Z}}\pmb{\nu}\\
\frac{\pmb{\nu}^\top\tilde{\mathbf{Q}}^{\top}\tilde{\mathbf{Q}}\pmb{\nu}}{\pmb{\nu}^\top\pmb{\nu}}&\geq c_1\frac{\pmb{\nu}^\top\tilde{\mathbf{Z}}^{\top}\tilde{\mathbf{Z}}\pmb{\nu}}{\pmb{\nu}^\top\pmb{\nu}}\\
    \min_{1\leq l\leq  k}\lambda_l\left(\tilde{\mathbf{Q}}^{\top}\tilde{\mathbf{Q}}\right)&\geq c_1 \min_{1\leq l\leq  k}\lambda_l\left(\tilde{\mathbf{Z}}^{\top}\tilde{\mathbf{Z}}\right)
\end{align*}
Since $\mathbf{P}$ is a projection matrix, then $\|\mathbf{P}\mathbf{y}\|^2\leq \|\mathbf{y}\|^2$, and then $0\leq c_1\leq 1$. Equation \eqref{trace1} can be expressed as:

\begin{align*}
    \mbox{tr}\left(\mathbf{M}\mbox{Var}(\hat{\pmb{\beta}})\mathbf{M}^\top\right)&\leq \frac{m_\rho}{ns_1}\sum_{l=1}^k \frac{\lambda_{\max}\left(\tilde{\mathbf{Z}}^\top \tilde{\mathbf{Z}}\right)}{c_1\lambda_{\min}\left(\tilde{\mathbf{Z}}^\top \tilde{\mathbf{Z}}\right)}\\
    &= \frac{km_\rho}{nc_1s_1} \frac{\lambda_{\max}\left(\tilde{\mathbf{Z}}^\top \tilde{\mathbf{Z}}\right)}{\lambda_{\min}\left(\tilde{\mathbf{Z}}^\top \tilde{\mathbf{Z}}\right)} \\
    &=\frac{km_\rho}{nc_1s_1} \frac{\lambda_{\max}\left(\mathbf{Z}^\top(\mathbf{A}^{-1})^\top \mathbf{A}^{-1}\mathbf{Z}\right)}{\lambda_{\min}\left(\mathbf{Z}^\top(\mathbf{A}^{-1})^\top \mathbf{A}^{-1}\mathbf{Z}\right)}\\
    &=\frac{km_\rho}{nc_1s_1} \frac{\lambda_{\max}\left(\mathbf{Z}^\top(\mathbf{A}\mathbf{A}^\top)^{-1}\mathbf{Z}\right)}{\lambda_{\min}\left(\mathbf{Z}^\top(\mathbf{A}\mathbf{A}^\top)^{-1}\mathbf{Z}\right)}
\end{align*}
Since $\mathbf{W}$ is stochastic matrix, let $c_w=\|\mathbf{W}\|_1$, the largest column sum of $\mathbf{W}$. We obtain that
\begin{align}
    \|\mathbf{W}\|_2\leq \sqrt{\|\mathbf{W}\|_1\|\mathbf{W}\|_\infty}& = \sqrt{c_w\cdot 1}=\sqrt{c_w}\nonumber\\
    \frac{1}{(1+|\rho|\sqrt{c_w})^2} &\leq \lambda_l\left((\mathbf{A}\mathbf{A}^\top)^{-1}\right)\leq \frac{1}{(1-|\rho|\sqrt{c_w})^2}\nonumber\\
    \mbox{tr}\left(\mathbf{M}\mbox{Var}(\hat{\pmb{\beta}})\mathbf{M}^\top\right)&\leq \frac{km_\rho\kappa^2_{\mathbf{A}}}{nc_1s_1} \frac{\lambda_{\max}\left(\mathbf{Z}^\top\mathbf{Z}\right)}{\lambda_{\min}\left(\mathbf{Z}^\top\mathbf{Z}\right)}=\frac{km_\rho\kappa^2_{\mathbf{A}}\kappa^2_{\mathbf{Z}}}{nc_1s_1}\label{eccote}
\end{align}
where  $$\kappa^2_{\mathbf{A}}\leq \left(\frac{1+|\rho|\sqrt{c_w}}{1-|\rho|\sqrt{c_w}}\right)^2\, \text{and }\kappa^2_{\mathbf{Z}}=\frac{\lambda_{\max}\left(\mathbf{Z}^\top\mathbf{Z}\right)}{\lambda_{\min}\left(\mathbf{Z}^\top\mathbf{Z}\right)}.$$ In this case, after applying a Principal Component Analysis (PCA) so that 
$\mathbf{Z}^\top\mathbf{Z}=\mathbb{I}_k$, 
we have $\kappa^2_{\mathbf{Z}}=1$, meaning that all eigenvalues are equal to one and 
the effect of ${\mathbf{Z}}$ over \eqref{eccote} is isolated, and becomes to:
\begin{align}
    \mbox{tr}\left(\mathbf{M}\mbox{Var}(\hat{\pmb{\beta}})\mathbf{M}^\top\right)&\leq \frac{km_\rho\kappa^2_{\mathbf{A}}}{nc_1s_1}\label{eccote2}
\end{align}
Is is clear that Equation \eqref{eccote2} increasing when $|\rho|$ tends to 1 and $|\phi_1|$ tends to 1. 

Since $\hat{\mathbf{Y}}$ is the BLUP \cite[pgs 269-274]{searle2009variance}, then $$\mathrm{MSE}(\hat{\mathbf{Y}}^{1}\mid\mathbf{Y}_a) - \mathrm{MSE}(\pmb{\mathbf{Y}}\mid\mathbf{Y}_a)\geq0,$$  and the improvement achieved by using the conditional predictor instead of the naive one can be expressed as
\begin{align*}
\Delta &= \mathbb{E}(\mathrm{MSE}(\hat{\mathbf{Y}}^{1}, \mathbf{Y})\mid \mathbf{Y}_a) -\mathbb{E}(\mathrm{MSE}(\hat{\mathbf{Y}}, \mathbf{Y})\mid \mathbf{Y}_a)\\
&=\|(\mathbb{I}_{nT}-\mathbf{P})\pmb{\mu}_{1a}\|^{2}
+ \operatorname{tr}\big((\mathbb{I}_{nT}-\mathbf{P})(\mathbb{I}_{nT}-\mathbf{P}_{1})\mathbf{B}\big)-\operatorname{tr}\big((\mathbb{I}_{nT}-\mathbf{P}_{1})\mathbf{B}\big)\\
&=\pmb{\mu}_{1a}^\top(\mathbb{I}_{nT}-\mathbf{P})\pmb{\mu}_{1a}
+ \operatorname{tr}\big((\mathbb{I}_{nT}-\mathbf{P})(\mathbb{I}_{nT}-\mathbf{P}_{1})\mathbf{B}\big)-\operatorname{tr}\big((\mathbb{I}_{nT}-\mathbf{P}_{1})\mathbf{B}\big)\\
&=\operatorname{tr}(\pmb{\mu}_{1a}^\top(\mathbb{I}_{nT}-\mathbf{P})\pmb{\mu}_{1a})
+ \operatorname{tr}\big((\mathbb{I}_{nT}-\mathbf{P})(\mathbb{I}_{nT}-\mathbf{P}_{1})\mathbf{B}-(\mathbb{I}_{nT}-\mathbf{P}_{1})\mathbf{B}\big)\\
&=\operatorname{tr}\big((\mathbb{I}_{nT}-\mathbf{P})\pmb{\mu}_{1a}\pmb{\mu}_{1a}^{\top}-\mathbf{P}(\mathbb{I}_{nT}-\mathbf{P}_{1})\mathbf{B}\big)
\end{align*}
And using Equations \eqref{Kron1}--\eqref{Kronn}, it is true that,
\begin{align}
    \Delta &= \pmb{\mu}^\top(\mathbb{I}_{nT}-\mathbf{P})\pmb{\mu}+\nonumber\\
    &
    \mbox{tr}\left((\mathbb{I}_{nT}-\mathbf{P})(\mathbb{I}_T\otimes\mathbf{D}_\rho)(\mathbf{Y}_a-\pmb{\mu}_a)(\mathbf{Y}_a-\pmb{\mu}_a)^\top(\mathbb{I}_T\otimes\mathbf{D}_\rho^\top)\right)+\nonumber\\
    &2\pmb{\mu}^\top(\mathbb{I}_{nT}-\mathbf{P})(\mathbb{I}_T\otimes\mathbf{D}_\rho^\top)(\mathbf{Y}_a-\pmb{\mu}_a)-\nonumber\\
    &\mbox{tr}\left(\pmb{\Sigma}_\mathbf{U}\otimes\left(\frac{1}{n}\mathbf{J}_n-\frac{1}{nm_\rho}\mathbf{J}_n\mathbf{F}^{-1}(\mathbf{F}^{-1})^{\top}\mathbf{J}_n\right)\mathbf{F}^{-1}(\mathbf{F}^{-1})^{\top}\right)\label{deltadelta}
\end{align}

The second term of Equation \eqref{deltadelta} can be expressed as:
\begin{align*}
    &\mbox{tr}\left((\mathbb{I}_{nT}-\mathbf{P})(\mathbb{I}_T\otimes\mathbf{D}_\rho)(\mathbf{Y}_a-\pmb{\mu}_a)(\mathbf{Y}_a-\pmb{\mu}_a)^\top(\mathbb{I}_T\otimes\mathbf{D}_\rho^\top)\right)\\
&=\mbox{tr}\left(\mathbf{m}^\top(\mathbb{I}_T\otimes\mathbf{D}_\rho^\top)\left(\mathbb{I}_{T}\otimes\left(\mathbb{I}_n -\frac{1}{n} \mathbf{J}_n\right)\right)(\mathbb{I}_T\otimes\mathbf{D}_\rho) \mathbf{m}\right)\\
&=\mbox{tr}\left(\mathbf{m}^\top \left(\mathbb{I}_{T}\otimes \mathbf{D}_\rho^\top\left(\mathbb{I}_n -\frac{1}{n} \mathbf{J}_n\right)\mathbf{D}_\rho\right) \mathbf{m}\right)\\
&=\mbox{tr}\left(\mathbf{m}^\top (\mathbb{I}_{T}\otimes k_\rho)\mathbf{m}\right)\\
&=k_\rho \mathbf{m}^\top \mathbf{m}
\end{align*}
when $k_\rho= \mathbf{D}_\rho^\top\left(\mathbb{I}_n -\frac{1}{n} \mathbf{J}_n\right)\mathbf{D}_\rho\geq 0$. The vector $\mathbf{D}_\rho=\mathbf{F}^{-1}(\mathbf{F}^{-1})^\top m_\rho^{-1}\mathbf{1}_n$ depends on the spatial autoregressive parameter $\rho$ through $\mathbf{F}=\mathbb{I}_n-\rho\mathbf{W}$. 
When $\rho=0$, $\mathbf{F}=\mathbb{I}_n$ and hence $\mathbf{D}_\rho = m_\rho^{-1}\mathbf{1}_n=\frac{1}{n}\mathbf{1}_n$, that is, $\mathbf{D}_\rho$ is proportional to the unit vector.
The third term of Equation \eqref{deltadelta} can be expressed as:
\begin{align*}
    &2\pmb{\mu}^\top(\mathbb{I}_{nT}-\mathbf{P})(\mathbb{I}_T\otimes\mathbf{D}_\rho)(\mathbf{Y}_a-\pmb{\mu}_a)\\
&=2\pmb{\mu}^\top\left(\mathbb{I}_T\otimes\left(\mathbb{I}_n-\frac{1}{n}\mathbf{J}_n\right)\mathbf{D}_\rho\right)(\mathbf{Y}_a-\pmb{\mu}_a)\\
&=2\pmb{\mu}^\top\left(\mathbb{I}_T\otimes\left(\mathbf{D}_\rho-\frac{1}{n}\mathbf{1}_n\right)\right)(\mathbf{Y}_a-\pmb{\mu}_a)
\end{align*}
And using the matrix form, it is clear that: 
\begin{align}
&=2(\pmb{\mu}_1^\top, \ldots, \pmb{\mu}_T^\top)\begin{pmatrix}
    \mathbf{D}_\rho-\frac{1}{n}\mathbf{1}_n & \mathbf{0} &\ldots & \mathbf{0}\\
     \mathbf{0} &\mathbf{D}_\rho-\frac{1}{n}\mathbf{1}_n &\ldots & \mathbf{0}\\
      \vdots &\vdots&\ddots & \vdots\\
       \mathbf{0} &\mathbf{0}&\ldots & \mathbf{D}_\rho-\frac{1}{n}\mathbf{1}_n \\
\end{pmatrix} \begin{pmatrix}
    Y_{a1}-\mu_{a1}\\
    Y_{a2}-\mu_{a2}\\
    \vdots\\
    Y_{aT}-\mu_{aT}\\
\end{pmatrix}\nonumber\\
&=2\sum_{t=1}^T\pmb{\mu}^\top_t\left(\mathbf{D}_\rho-\frac{1}{n}\mathbf{1}_n\right) (Y_{at}-\mu_{at})\nonumber\\
&=2\left(\sum_{t=1}^T\pmb{\mu}^\top_t (Y_{at}-\mu_{at})\right)\left(\mathbf{D}_\rho-\frac{1}{n}\mathbf{1}_n\right)\nonumber\\
&=2\begin{pmatrix}
    \sum_{t=1}^T{\mu_{1t}} (Y_{at}-\mu_{at})\\
    \sum_{t=1}^T{\mu_{2t}} (Y_{at}-\mu_{at})\\
    \vdots\\
    \sum_{t=1}^T{\mu_{nt}} (Y_{at}-\mu_{at})\\
\end{pmatrix}^\top\left(\mathbf{D}_\rho-\frac{1}{n}\mathbf{1}_n\right)\nonumber\\
&=\tilde{\pmb{\mu}}^\top \tilde{\pmb{\rho}}\label{ecrhotilde}
\end{align}
when $\pmb{\mu}_{t}^\top= (\mu_{1t}, \ldots, \mu_{nt})$, $\tilde{\pmb{\mu}}^\top=\left(\sum_{t=1}^T{\mu_{1t}} (Y_{at}-\mu_{at}), \ldots, \sum_{t=1}^T{\mu_{nt}} (Y_{at}-\mu_{at})\right)$, $\tilde{\pmb{\rho}}=\left(\mathbf{D}_\rho-\frac{1}{n}\mathbf{1}_n\right)$, and $(\mathbf{Y}_a-\pmb{\mu}_a)^\top=(Y_{a1}-\mu_{a1}, \ldots, Y_{aT}-\mu_{aT})$. Additionally, it is clear that $\tilde{\pmb{\rho}}^\top\mathbf{1}_n=0$. $\tilde{\pmb{\mu}}^\top \tilde{\pmb{\rho}}$ has expected value $0$ because $\mathbb{E}(\mathbf{Y}_a-\pmb{\mu}_a)=0$. If ${\rho}=0$, then the value is $0$, because $\mathbf{D}_\rho=\frac{1}{n}\mathbf{1}_n$.
The fourth term of Equation \eqref{deltadelta} can be expressed as:
\begin{align*}
    &\mbox{tr}\left(\pmb{\Sigma}_\mathbf{U}\otimes\left(\frac{1}{n}\mathbf{J}_n-\frac{1}{nm_\rho}\mathbf{J}_n\mathbf{F}^{-1}(\mathbf{F}^{-1})^{\top}\mathbf{J}_n\right)\mathbf{F}^{-1}(\mathbf{F}^{-1})^{\top}\right)\\
    &=\mbox{tr}\left(\pmb{\Sigma}_\mathbf{U}\right)\mbox{tr}\left(\left(\frac{1}{n}\mathbf{J}_n-\frac{1}{nm_\rho}\mathbf{J}_n\mathbf{F}^{-1}(\mathbf{F}^{-1})^{\top}\mathbf{J}_n\right)\mathbf{F}^{-1}(\mathbf{F}^{-1})^{\top}\right)\\
    &=\mbox{tr}\left(\pmb{\Sigma}_\mathbf{U}\right)\mbox{tr}\left(\left(\frac{1}{n}\mathbf{J}_n-\frac{1}{n m_\rho}\mathbf{1}_n\mathbf{1}_n^\top\mathbf{F}^{-1}(\mathbf{F}^{-1})^{\top}\mathbf{1}_n\mathbf{1}_n^\top\right)\mathbf{F}^{-1}(\mathbf{F}^{-1})^{\top}\right)\\
    &=\mbox{tr}\left(\pmb{\Sigma}_\mathbf{U}\right)\mbox{tr}\left(\left(\frac{1}{n}\mathbf{J}_n-\frac{m_\rho}{nm_\rho}\mathbf{1}_n\mathbf{1}_n^\top\right)\mathbf{F}^{-1}(\mathbf{F}^{-1})^{\top}\right)\\
    &=\mbox{tr}\left(\pmb{\Sigma}_\mathbf{U}\right)\mbox{tr}\left(\left(\frac{1}{n}\mathbf{J}_n-\frac{1}{n}\mathbf{J}_n\right)\mathbf{F}^{-1}(\mathbf{F}^{-1})^{\top}\right)\\
    &=\mathbf{0}
\end{align*}

\end{proof}

\end{document}